\documentclass[11pt]{article}
\pdfoutput=1
\usepackage{graphicx}
\usepackage{xcolor}
\usepackage{comment}
\usepackage{adjustbox} 
\usepackage{appendix}
\usepackage{latexsym,amsmath,amsfonts,amssymb,booktabs}
\usepackage[font=small]{caption}
\usepackage{slashed,upgreek,amscd,cancel,tensor,color}
\usepackage{adjustbox}
\usepackage{soul}
\usepackage[normalem]{ulem}
\usepackage[numbers,compress,square]{natbib}
\usepackage{epsfig,latexsym}
\usepackage{amsmath}
\usepackage{enumerate}  
\usepackage{tensor}
\usepackage{braket}
\usepackage{tikz}
\usepackage{tikz-feynman}
\usepackage{booktabs}
\usepackage{multirow}
\usepackage{siunitx}
\tikzfeynmanset{compat=1.1.0}
\usepackage[pdfencoding=auto]{hyperref} 
\usepackage{url}
\numberwithin{equation}{section}
\usepackage{doi}
\usepackage{subcaption}
\definecolor{MyBlue}{rgb}{0.15,0.15,0.70}
\definecolor{linkblue}{rgb}{0,0,0.8}
\definecolor{linkgreen}{rgb}{0,0.5,0}

\hypersetup{
colorlinks=true,
citecolor=linkgreen,
linkcolor=linkblue,
urlcolor=linkblue
}

\input epsf
\usepackage{amssymb}
\usepackage{amsmath}
\usepackage{amsfonts}
\usepackage{upgreek}
\usepackage{latexsym}
\usepackage{stfloats}
\usepackage{afterpage}

\newcommand{\hinvMpc}{\,h\, {\rm Mpc}^{-1}\,}

\usepackage{xspace}

\newcommand{\JAX}{\texttt{JAX}\xspace}
\newcommand{\PyBird}{\texttt{PyBird}\xspace}
\newcommand{\JAXBird}{\texttt{PyBird-JAX}\xspace}

\begin{document}
\def\thefootnote{\fnsymbol{footnote}}
\setcounter{page}{1} \baselineskip=15.5pt \thispagestyle{empty}

\vspace*{-25mm}

\vspace{0.5cm}

\begin{center}

{\Large \bf Debiasing inference in large-scale structure \\[0.2cm]
with non-flat volume measures
}  \\[0.7cm]
{\large   Alexander Reeves${}^{1}$,  Pierre Zhang${}^{1,2,3}$, and Henry Zheng${}^{4,5}$\\[0.7cm]}

\end{center}

\begin{center}

\vspace{.0cm}

\begin{small}

{ \textit{  $^{1}$ Institute for Particle Physics and Astrophysics, ETH Z\"urich, 8093 Z\"urich, Switzerland}}
\vspace{.05in}

{ \textit{  $^{2}$ Institut f\"ur Theoretische Physik, ETH Z\"urich, 8093 Z\"urich, Switzerland}}
\vspace{.05in}

{ \textit{ $^{3}$ Dipartimento di Fisica “Aldo Pontremoli”, Universit\`a degli Studi di Milano, 20133 Milan, Italy}}

{ \textit{  $^{4}$ Stanford Institute for Theoretical Physics, Physics Department, Stanford University, Stanford, CA 94306}}
\vspace{.05in}

{ \textit{  $^{5}$ Kavli Institute for Particle Astrophysics and Cosmology, SLAC and Stanford University, Menlo Park, CA 94025}}
\vspace{.05in}

\end{small}
\end{center}

\vspace{0.5cm}

\begin{abstract}
Increasingly large parameter spaces, used to more accurately model precision observables in physics, can paradoxically lead to large deviations in the inferred parameters of interest --- a bias known as \emph{volume projection effects} --- when marginalising over many nuisance parameters. 
For posterior distributions that admit a Laplace expansion, we show that this artefact of Bayesian inference can be mitigated by defining expectation values with respect to a non-flat volume measure, such that the posterior mean becomes unbiased on average.  
We begin by finding a measure that ensures the mean is an unbiased estimator of the mode. 
Although the mode itself, as we rediscover, is biased under sample averaging, this choice yields the least biased estimator due to a cancellation we clarify. 
We further explain why bias in marginal posteriors can appear relatively large, yet remains correctable, when the number of nuisances is large. 
To demonstrate our approach, we present mock analyses in large-scale structure (LSS) wherein cosmological parameters are subject to large projection effects (at the 1-2$\sigma$ level) under a flat measure, that are however recovered at high fidelity ($<0.1\sigma$) when estimated using non-flat counterparts. 
Our cosmological analyses are enabled by \texttt{PyBird-JAX}, a fast, differentiable pipeline for LSS developed in our companion paper~\cite{paper1}.
\end{abstract}

\newpage

\tableofcontents

\vspace{.5cm}

\def\thefootnote{\arabic{footnote}}
\setcounter{footnote}{0}

%
%
%
%
\pagebreak 

\section{Introduction}  \label{sec:intro}

When the overall signal-to-noise ratio is high, there is a priori no difficulty in building confidence in our physical results. 
Probability distributions of inferred parameters entering our physical models generally converge to Gaussians with variance given by the inverse Fisher information in the large data limit. Moreover, if the estimator is unbiased, the true parameter values are recovered asymptotically. 
We are however not always in this ideal situation. 
Sometimes the data is weak, yet we would like to make statements about some model.
When the model is a linear function of the parameters, as it is often the case near equilibrium, the resulting likelihoods are Gaussian, even when data is limited. 
In this scenario, credible intervals remain well-defined and unambiguous. 
Complications arise when it becomes necessary to consider models with nonlinear dependence. 
Suppose we are provided with data spanning a wide range of scales accessible via an effective field theory approach.  
To extract the maximum amount of information, we naturally aim to compute predictions at the highest available order, thereby extending our reach to smaller scales. 
Each successive order introduce additional effective parameters (typically increasing in number with the perturbative order) which must be marginalised over in the inference, contributing to the overall noise. In such situations where we are adding extra parameters that couple with those of interest, we might be facing large departure from Gaussianity in the posterior distributions of our model parameters.

Non-Gaussianities in posteriors are not undesirable per se. 
After all, they participate in weighting faithfully the most credible region in the multidimensional parameter space corresponding to the assumed underlying model of the observations we are given. 
However, except when assessing the consistency of two datasets under a specific model, or in model comparison for a given dataset, we are never really interested in the full $N$-dimensional posterior. 
Instead, we seek to understand the probability on a subset of parameters of interest such that we would marginalise over the others, that we call nuisances. 
In general, projecting out the extra dimensions and taking expectation values on the resulting 1D posteriors can lead to significant shifts in their mean with respect to the truth. 
The resulting credible intervals are said to suffer from \emph{volume projection effects}. 
While there is a substantial body of literature on mitigating such biases in parameter inference, dating back to the seminal work of Welch and Peers~\cite{Welchs,peers}, a comprehensive review is beyond the scope of the present paper. 
We instead direct readers to the excellent reviews on probability matching priors by Reid, Mukerjee, and Fraser~\cite{Reid}, as well as the foundational contributions on reference priors in objective Bayesian inference by Berger, Bernardo, and Sun~\cite{Berger}. 
For a perspective more grounded in the physics community, see also ref.~\cite{DAgostini:1999gfj}. 
Let us now elaborate on an actual situation we are facing. 

In cosmology, the complicated galaxy formation physics that we integrate out by looking at the cosmic maps at sufficiently long distances, is captured into a set of effective parameters~\cite{Fry:1992vr,Heavens:1998es,McDonald:2006hf,McDonald:2009dh,Baumann:2010tm,Senatore:2014eva}. 
These parameters may be of use to understand certain astrophysical processes. However they are not the main target as their link to the fundamental properties of the Universe is far from understood. 
With no extra a priori information, these nuisance parameters are usually marginalised over with large, order-of-magnitude, priors, estimated from naturalness considerations~\cite{DAmico:2019fhj,DAmico:2022osl}. 
Recent studies have shown that when the data volume is small relative to the large volume of nuisance parameters that must be marginalised over, this marginalisation can lead to significant volume projection effects in the inferred cosmological parameters~\cite{Ivanov:2019pdj,Simon:2022lde,DESI:2024jis}. 
One attitude is to say that this is not a real problem since this is just an artefact that comes from the way we define, arbitrarily, our expectations. 
Yet, if we want to know how much the Universe is expanding today, ultimately we want to make a statement about a handful of numbers, the Hubble constant, the abundance of matter, etc., for which we assume there exists a true, physical, value (until the data reject the model). 
The practical question is then to find a consensus to report inferred values in a definite way, especially when one wants to make claim about potential tensions that could signal departure from the standard model. 
Here a definite way means that the numbers we quote should be as close as possible to the nominal truth assuming our model contains all that is necessary to describe the data. 
This is a question of robustness of our statistical inference, which is well defined as we will argue in this paper.
Usually, moments of posterior distributions are taken upon integration over the parameter space volume with Lebesgue measure. 
However, there is no fundamental argument to consider that the manifold defined by the parameter space is Euclidian other than simplicity. 
Perhaps as a guiding principle we would like that the volume measure respect desirable symmetry properties of the manifold living in the parameter space defined by the posterior distribution. 
Ultimately, a `good' measure should be a measure that allows us to define `good' statistics in the sense of their usual desired properties: unbiasedness, efficiency, etc.\footnote{Strictly, bias and efficiency (in the sense of achieving the Cramér-Rao lower bound) are frequentist concepts that do not directly apply for Bayesian inference. 
In this paper, we are taking the view that the data can be repeatedly sampled from fixed, true parameter values. 
Although we observe only one sky, we can observe many independent regions that each contain a large number of spatial modes. 
Each region is then immune to cosmic variance, and is considered to be a representative sample of the whole Universe. 
This is nothing but a re-statement of the \emph{fair sample hypothesis} of J. Peebles. 
More on this in sec.~\ref{sec:asymptotic}. 
}  
 
From this standpoint, it seems that defining appropriately a notion of `geometry' in our inference should allow us to reconcile our expectation values with the nominal truth in an average sense, at least locally and under certain smoothness conditions on the parameter space. 
The rest of the paper is dedicated to clarify what this means.  
In general, we will be able to pick a statistical geometry that increases the efficiency of our parameter estimation asymptotically for close to normal distributions in a sense that we will precise. 
See ref.~\cite{Kass0} for an introduction to curved geometries in statistical inference with an angle close to the one we are taking here. 

Our main result is a proof that there exists an appropriate choice of integration measure, defining expectation values taken over distributions that admit a Laplace expansion, such that marginal posterior means are unbiased (at leading order in the asymptotic expansion) when averaged over data samples. 
Along the way, we will re-discover that the maximum likelihood estimator (MLE), 
\begin{equation}\label{eq:mle}
\text{MLE}(\pmb\theta) := \arg \max_{\pmb\theta} \mathcal{L}(y|\pmb\theta) \ ,
\end{equation}
where $\mathcal{L}$ is some likelihood of data $y$, 
is biased on average over noisy data samples~\cite{McCullagh}. 

Our paper is organised as follows. 
In sec.~\ref{sec:asymptotic}, we first study a simple example to guide our understanding on how volume projection effects can lead to bias in parameter estimation, then define our inference, and finally present a posterior expansion based on the Laplace method. 
Next, we underpin, under sample average, the bias in the posterior mean in sec.~\ref{sec:mean} and in the posterior mode in sec.~\ref{sec:mode}.
After revisiting the posterior mean in light of the mode bias in sec.~\ref{sec:mean_revisited}, we then show in sec.~\ref{sec:estimator} that the leading average bias in the mean can be removed by choosing an appropriate volume measure when defining our expectation values. 
We elucidate why volume projection effects can appear relatively large, yet remain correctable, in sec.~\ref{sec:Nbias}. 
We then turn to concrete examples in cosmological large-scale structure (LSS) analyses in sec.~\ref{sec:lss}, using our differentiable inference pipeline developed in our companion paper~\cite{paper1}. 
We conclude in sec.~\ref{sec:conclusions}. 
Additional materials and details are relegated to the appendices.

\paragraph{Notations and conventions}
We use the Einstein summation convention where repeated indices are summed over, \textit{i.e.}, $u_\mu v_\mu = u^\mu v_\mu = \sum_{\mu}u_\mu v_\mu$. 
Furthermore, we introduce the notation $\mathcal{O}_{i_1\dots i_n;\mu_1\dots \mu_m} = \partial_{\mu_1 \dots \mu_m} \mathcal{O}_{i_1\dots i_n}$, where partial derivatives $\partial_{\mu_1 \dots \mu_m}$ are taken with respect to the model parameters $\pmb\theta$ introduced below.

\section{Inference, asymptotically}\label{sec:asymptotic}

In this section, we start by setting the stage by defining how we usually infer model parameters of interest from given observations. 
After presenting a simple toy model to build understanding on how the volume measure projects onto marginal posterior distributions in sec.~\ref{sec:toy}, we then turn our attention to general smooth distributions that converge asymptotically to normal distributions in sec.~\ref{sec:large-n}, presenting their Laplace expansion in sec.~\ref{sec:laplace}.

\paragraph{Setup}

Let $ \pmb{\theta} \in \mathbb{R}^N $ denote the parameters of a model $ m(\pmb{\theta}) $, and let $ y \in \mathbb{R}^d$ be the observed data. We assume the model is \emph{well specified}, meaning that the data $ y $ is generated according to some true parameter values $ \pmb{\theta_\dagger} \in \mathbb{R}^N$.
We define the \emph{Gaussian likelihood} function $ \mathcal{L}(y | \pmb{\theta}) $ such as
\begin{equation}\label{eq:likelihood}
-2 \log \mathcal{L}(y | \pmb{\theta}) = (m(\pmb{\theta}) - y)^T C^{-1} (m( \pmb{\theta}) - y) \ ,
\end{equation}
where we have dropped from the notation a normalisation factor. 
Here $ C\in \mathbb{R}^{d \times d}$ is a known covariance matrix of the observational noise. 
Given a function $X(y)$, we define the expectation value of $X$ under the likelihood induced by the true parameter $ \pmb{\theta_\dagger}$ as
\begin{equation}\label{eq:average}
\braket{X} \equiv \int dy \ \mathcal{L}(y|\pmb{\theta_\dagger}) \ X \ .
\end{equation}
This represents the sample (ensemble) average over data realisations $y \sim \mathcal{L}(y|\pmb{\theta_\dagger})$, assuming the model correctly describes the data-generating process. 
As such, $\braket{y} = m(\pmb{\theta_\dagger}) \equiv m_\dagger$ and $\braket{yy^T}- \braket{y}\braket{y}^T=C$. 
Given a prior $\pi(\pmb{\theta})$, we can then infer the posterior distribution $\mathcal{P}(\pmb{\theta} | y) \propto  \mathcal{L}(y | \pmb{\theta}) \pi(\pmb{\theta})$ following Bayes theorem. 
We consider a large flat prior except when stated otherwise. 
For clarity, we will refer to eq.~\eqref{eq:average} as \textit{sample averaging} and reserve the terminology \textit{expectation value} to \textit{expectation value under the posterior distribution} that we introduce now. 

\paragraph{Expectation values}
Given a posterior distribution $\mathcal{P}$ of inferred parameters $\pmb{\theta}$, we define $p$-moments as expectation values over $\theta_{\alpha_{1 \dots  p}} \equiv \theta_{\alpha_1} \times \dots \times \theta_{\alpha_p}$ weighted by the distribution, 
\begin{equation}\label{eq:moment}
\mathbb{E}_{\mathcal{P},\mathcal{M}}[\theta_{\alpha_{1 \dots  p}}] = \frac{1}{\mathcal{Z}_{\mathcal{P},\mathcal{M}}} \tilde{\mathbb{E}}_{\mathcal{P},\mathcal{M}}[\theta_{\alpha_{1 \dots  p}}] \ , \qquad \tilde{\mathbb{E}}_{\mathcal{P},\mathcal{M}}[\theta_{\alpha_{1 \dots  p}}] = \int  \mathcal{M(\pmb{\theta}}) \ \theta_{\alpha_{1 \dots  p}} \ \mathcal{P}(\pmb{\theta} | y) \ , 
\end{equation}
where $\mathcal{M(\pmb{\theta}})$ is the integration measure and $\mathcal{Z}_{\mathcal{P},\mathcal{M}} \equiv \tilde{\mathbb{E}}_{\mathcal{P},\mathcal{M}}[1]$ is the normalising evidence. 
When choosing Lebesgue measure, $\mathcal{M} \equiv d^N \pmb{\theta}$, we simply denote the $p$-moments as $\mathbb{E}_{\mathcal{P}}[\theta_{\alpha_{1 \dots  p}}]$. 
In particular, the credible interval for parameter $\theta_\alpha$ reads $\mu_\alpha \pm \sigma_\alpha$, where $\mu_\alpha$ is the mean and $\sigma_\alpha$ the standard deviation defined as
 \begin{equation}
\mu_\alpha = \mathbb{E}_{\mathcal{P},\mathcal{M}}[\theta_\alpha] \ , \qquad \sigma_\alpha^2 = \mathbb{E}_{\mathcal{P},\mathcal{M}}[\theta_\alpha^2] - \mu_\alpha^2 \ .
 \end{equation}
In this paper, we will mainly focus on the mean (1-moment) estimator. 

\subsection{Biased expectations: a toy model}\label{sec:toy}
Before delving into general posterior distributions (at least of the kind we usually encounter in physics), it is instructive to look at a simple two-parameter example where marginalising over one deemed nuisance leads to bias on the resulting posterior for the other one of interest. 
Let us consider the following model for a scalar observation (data vector of size one) depending on two parameters $\theta_0$ and $\theta_1$, 
\begin{equation}\label{eq:toy}
m(\theta_0, \theta_1) = \theta_0 + \theta_1 + \alpha \theta_0 \theta_1 \ , 
\end{equation}
where $\alpha$ is a small coupling constant ($\alpha \ll 1$). 
Given noiseless synthetic data and assuming that the true parameter values are $\theta_0^\dagger = 0,\theta _1^\dagger = 0$ ($y\equiv0$ accordingly), the resulting posterior distribution is given by, up to an irrelevant multiplicative factor accounting for the data covariance,
\begin{equation}
-2 \log \mathcal{P}(\theta_0, \theta_1)  = m(\theta_0, \theta_1)^2 + \theta_0^2 \ ,
\end{equation}
where we have further imposed a Gaussian prior on $\theta_0$ with unit variance centred on $0$.

\paragraph{Volume projection effects} Marginalising over $\theta_1$ yields the marginal posterior distribution of $\theta_0$,
\begin{equation}
\mathcal{P}(\theta_0) = \int d\theta_1 \, \mathcal{P}(\theta_0, \theta_1) \propto \frac{1}{1+\alpha\theta_0} \, \exp \left({-\frac{1}{2}\theta_0^2} \right) \simeq  \exp \left({-\frac{1}{2}\theta_0^2} \right) \, \left(1-\alpha\theta_0 + \mathcal{O}(\alpha^2)\right) \ , 
\end{equation}
where we have used the formula for Gaussian integrals in the third equality (dropping an irrelevant $(2\pi)^{1/2}$-factor) and expanded in small $\alpha$ in the last equality. 
Defining the Gaussian generating functional 
\begin{equation}\label{eq:ggf}
G_0[j] \equiv \int d\theta_0 \, \exp \left({-\frac{1}{2}\theta_0^2 + j \theta_0}\right) \propto \exp \left(\frac{j^2}{2} \right) \ ,
\end{equation}
the evidence can be computed as
\begin{equation}
\mathcal{Z}_\mathcal{P} = \int d\theta_0 \, \mathcal{P}(\theta_0) = G_0|_{j=0} - \alpha \left. \frac{\partial G_0}{\partial j}\right|_{j=0} = G_0|_{j=0} \ ,
\end{equation}
where we have used that $\partial G_0/\partial j|_{j=0} \propto j\exp(j^2/2)|_{j=0} = 0$. 
Then, the mean of $\theta_0$ is given by
\begin{equation}
\mathbb{E}_{\mathcal{P}}[\theta_0] = \frac{1}{\mathcal{Z}_\mathcal{P}} \int d\theta_0 \, \theta_0 \, \mathcal{P}(\theta_0) = \left. \frac{1}{G_0}\frac{\partial G_0}{\partial j}\right|_{j=0} - \alpha \left. \frac{1}{G_0}\frac{\partial^2 G_0}{\partial^2 j}\right|_{j=0} = - \alpha \ .
\end{equation}
Therefore, the bias in the mean of the marginal posterior of $\theta_0$ is proportional to $\alpha$, the coupling with the marginalised parameter $\theta_1$. 
It is easy to convince ourselves that, the bias in the variance, in contrast, starts at $\mathcal{O}(\alpha^2)$. 

\paragraph{Non-Euclidian measure}
Consider now defining the expectation values instead with volume measure $\mathcal{M}_\mathcal{F} = \sqrt{\det \mathcal{F}} d\theta_0 d\theta_1$, where the Fisher information matrix is defined as $\mathcal{F}_{\mu\nu} := - \partial_{\mu\nu} \log \mathcal{P}$. 
By expanding in small $\alpha$ the volume element $\sqrt{\det \mathcal{F}} \simeq 1-\alpha \theta_1$
and defining the Gaussian generating functional
\begin{equation}\label{eq:g1}
G_1[j] \equiv \int d\theta_1 \, \exp \left({-\frac{1}{2} \left(m(\theta_0,\theta_1)^2 + \theta_0^2\right)+ j \theta_1}\right) \ ,
\end{equation}
we can perform the integration over $d\theta_1$ as
\begin{equation}
\int d\theta_1 \, (1-\alpha \theta_1) \, \mathcal{P}(\theta_0, \theta_1) = G_1|_{j=0} - \alpha \left. \frac{\partial G_1}{\partial j}\right|_{j=0} \ ,
\end{equation}
yielding
\begin{equation}
\propto \exp \left({-\frac{1}{2}\theta_0^2} \right) \left( \frac{1}{1 + \alpha\theta_0} + \frac{\alpha\theta_0}{(1 + \alpha\theta_0)^2} \right) \simeq \exp \left({-\frac{1}{2}\theta_0^2} \right) \left(1-\alpha^2 \theta_0^2 + \mathcal{O}(\alpha^3)\right) \ .
\end{equation}
From there we can follow the same steps as above, and we see that in the small-$\alpha$ limit, the linear term in $\alpha$ now cancels. 
The linear bias in the mean of $\theta_0$ thus vanishes when defining expectation values with respect to the measure $\mathcal{M}_\mathcal{F}$.  
In the rest of the paper, we generalise the picture to arbitrary asymptotic normal distributions as defined below, also considering the presence of noise. 

\subsection{Power counting}\label{sec:large-n}
\paragraph{Decoupling limit} Our toy model can be generalised in a straightforward manner. 
For a set of $N$ parameters $\pmb{\theta} = \{\theta_1, \dots, \theta_N \}$, assuming the model $m$ to be smooth function of $\pmb{\theta}$, we can expand $m$ around the true parameter values $\pmb{\theta_\dagger}$ as
\begin{equation}\label{eq:toy2}
m(\pmb{\theta}) = m(\pmb{\theta_\dagger}) + \partial_\mu m|_{\pmb{\theta}=\pmb{\theta_\dagger}} (\theta_\mu - \theta_\mu^\dagger) +  \partial_{\mu\nu} m|_{\pmb{\theta}=\pmb{\theta_\dagger}} (\theta_\mu - \theta_\mu^\dagger)(\theta_\nu - \theta_\nu^\dagger) + \dots
\end{equation}
Here $\alpha_{\mu\nu} \equiv \partial_{\mu\nu} m|_{\pmb{\theta}=\pmb{\theta_\dagger}}$ play the role of the coupling constants. 
In the limit of $\alpha_{\mu\nu}\rightarrow 0$, the model becomes linear in $\pmb{\theta}$. 
The resulting posterior distribution is then Gaussian. 
Non-Gaussianities in the posterior thus follow departure from the decoupling limit. 
While instructive, this limit is somewhat artificial: the model choice is not dictated by a desired form of the posterior (\textit{e.g.}, one with unbiased mean and minimal variance). 
Instead, we have in principle one handle: repeat the experiment to gather more data. 

\paragraph{Large sample limit} 
Usually, eq.~\eqref{eq:average} is thought to be equivalent to averaging over $n$ independent samples obtained from repeating the experiment $n$ times (\textit{i.e.,} drawn from $\mathcal{L}$) and taking $n \rightarrow \infty$. 
In cosmology, we appeal to the ergodic theorem such that $\braket{\cdot}$ can be thought of as an average over $n$ spatial regions. 
For example, we can imagine observing $n$ uncorrelated patches of sky, each corresponding to a unit data volume $V$, such that the likelihood $\mathcal{L}$ is constructed over these $n$ independent observations. 
This situation corresponds to redefining $C \rightarrow n^{-1} \, C$, $(m-y) \rightarrow n^{-1/2}(m-y)$, as $C$ is inversely proportional to the data volume, and we remind that $y$ is drawn from a normal distribution with covariance $C$. 
Taking $n \rightarrow \infty$ then makes the large sample limit manifest. 
Consequently, the Fisher information matrix, being inversely proportional to $C$, scales as $\mathcal{F} \rightarrow n\mathcal{F}$. 
$\pmb{\delta} \equiv \pmb{\theta}-\pmb{\theta_\dagger}$ has approximately variance $\mathcal{F}^{-1}$, thus scales as $\pmb{\delta}\rightarrow n^{-1/2}\pmb{\delta}$. 
Plugging this scaling into eq.~\eqref{eq:toy2} shows that, in the limit of large $n \gg 1$, second-order terms become small compared to the linear terms. 
Posterior non-Gaussianities are thus suppressed in the large sample limit, akin to the decoupling limit. 
In the following, when studying the situation at finite $n$, we use $n$ as a bookkeeper for our posterior and moment expansions, eventually setting $n \equiv 1$ in the final results (and sometimes in the notation when convenient). 
In other words, we are simply making explicitly manifest in our equations the dependence on the number of trials (\textit{i.e.}, the number spatial patches in a cosmological context) encompassing the total dataset. Setting $n = 1$ amounts to simply reabsorbing the dependence into the covariance. 

\paragraph{Asymptotic normal distribution} We say that the posterior probability distribution $\mathcal{P}$ is \emph{asymptotic normal} if as $n \rightarrow \infty$, $\mathcal{P}(\pmb{\theta}|y)$ converges to a multivariate normal distribution $\mathcal{N}(\pmb{\theta_\dagger}, \mathcal{F}^{-1}_\dagger)$ centred on $\pmb{\theta_\dagger}$ with covariance given by the inverse Fisher information matrix $\mathcal{F}^{-1}_\dagger \equiv (\mathcal{F}|_{\pmb{\theta}=\pmb{\theta_\dagger}})^{-1}$. 
Starting from a Gaussian likelihood and assuming a regular model parametrisation, no model misspecification, uninformative priors, with the number of parameters $N$ remaining finite as $n \rightarrow \infty$, if $\mathcal{P}$ has a global maximum (\textit{i.e.}, is unimodal), then $\mathcal{P}$ is necessarily asymptotically normal by the Bernstein–von Mises theorem. 
This is the usual situation in physics where we have a parametric model with parameters representing physically meaningful observables, \textit{i.e.}, for which definite values can be measured out of the data, given enough data (assuming that the model is not misspecified). 
When $\mathcal{P}$ is asymptotically normal, it satisfies the conditions outlined in app.~\ref{app:validity} for applying the Laplace method, which we present next. 
In sec.~\ref{sec:efficient}, we assess the bias and the efficiency of the mean ($1$-moment) estimator defined in eq.~\eqref{eq:moment} for asymptotic normal distributions. 
Said differently, our goal is to understand how to construct an estimator for the mean that is the least biased, with minimal variance, at finite data size $n$. 

\subsection{Laplace expansion}\label{sec:laplace}

Let $\mathcal{P}$ be an asymptotically normal distribution. 
It has a global maximum $\pmb{\theta_*}$, otherwise called the posterior mode, is smooth around $\pmb{\theta_*}$, and 
decays exponentially in distant regions as stated in app.~\ref{app:validity}. 
Then, $\mathcal{P}$ can be expanded around $\pmb{\theta_*}$ using the Laplace method, which we now detail.\footnote{Since the mode $\pmb{\theta_*}$ is itself a random variable, one might be inclined to perform instead a straightforward Taylor expansion around the true parameter value $\pmb{\theta_\dagger}$.  
However, at finite sample size $n$ and at finite order in the expansion, this approach does not necessarily yield a convergent or well-defined expansion for the $p$-moments. 
For instance, $\pmb{\theta_\dagger}$ may be sitting at a local minimum (instead of being at a maximum). 
The expansion of $p$-moments is only well-defined around the global maximum $\pmb{\theta_*}$, where the distribution is locally peaked (see app.~\ref{app:validity}). 
Ultimately, inference bias must be evaluated relative to the nominal true value $\pmb{\theta_\dagger}$. 
For clarity, we will initially assume that the mode is unbiased in sec.~\ref{sec:mean}. 
This assumption will be relaxed in sec.~\ref{sec:mode}, and the resulting correction will be incorporated in sec.~\ref{sec:mean_revisited}. 
}

Locally, the posterior distribution $\mathcal{P}$ can be written as a Gaussian part perturbed with contributions coming from expanding around $\pmb{\theta_*}$.
Defining $n^{-1/2} \pmb{\delta} = \pmb{\theta} - \pmb{\theta_*}$ , 
\begin{align}
\mathcal{P}(\pmb{\theta}|\mathcal{F}, \pmb{j}) =  \mathcal{P}(\pmb{\theta_*}) & \exp \left( -\frac{1}{2}\delta_\mu \mathcal{F}_{\mu\nu} \delta_\nu + j_\mu \delta_\mu \right) \times 
\Bigg\{ 1 + \frac{n^{-1/2}}{2} \left[  j_{\mu;\nu}  \delta_\mu \delta_\nu - \frac{1}{2}\mathcal{F}_{\mu\nu;\rho} \delta_\mu \delta_\nu \delta_\rho \right]  \nonumber \\
&  + \frac{n^{-1}}{2} \left[ \frac{1}{3} j_{\mu;\nu\rho}  \delta_\mu \delta_\nu \delta_\rho - \frac{1}{4}\left(\frac{1}{2}\mathcal{F}_{\mu\nu;\rho\sigma} + \frac{1}{3} \mathcal{F}^\mu_{\nu;\rho\sigma} \right) \delta_\mu \delta_\nu \delta_\rho \delta_\sigma \right] \Bigg\} + \dots  \ , \label{eq:p_expand}
\end{align}
where $\mathcal{F}$, $\pmb j$, and their derivatives are evaluated at $\pmb{\theta_*}$. 
Here and in the rest of this section, $\dots$ refers to $\mathcal{O}\left(n^{-3/2}\right)$ when not specified. 
Why we choose to work at this order will become clear in the following. 
The Fisher information matrix and the source term are defined respectively as
\begin{align}\label{eq:fisher}
n \mathcal{F}_{\mu\nu}(\pmb{\theta}) & := \partial_\mu m(\pmb{\theta})^T C^{-1} \partial_\nu m(\pmb{\theta}) \  , \\
n^{1/2} j_\mu(\pmb{\theta}) & := \partial_\mu m(\pmb{\theta})^T C^{-1} \Delta_* \ , \qquad \Delta_* \equiv y- m( \pmb{\theta_*}) \ .\label{eq:source}
\end{align}
In particular, note that $j_{\mu;\nu \dots}(\pmb{\theta}) = \partial_{\mu\nu \dots}m(\pmb{\theta})^T C^{-1} \Delta_*$, since $\Delta_*$ does not depend explicitly on $\pmb{\theta}$. 
Similarly, we have also introduced a non-symmetric Fisher derivatives,
\begin{equation}\label{eq:nosymfish}
n \mathcal{F}_{\nu;\rho...}^\mu(\pmb{\theta})  := (\partial_\mu m|_{\pmb{\theta}=\pmb{\theta^*}})^T \, C^{-1}   \partial_{\nu\rho...} m(\pmb{\theta})\ , 
\end{equation}
which act only on the right $\pmb{\theta}$-dependent piece of the product while leaving the left piece untouched as evaluated on $\pmb{\theta^*}$ explicitly from the start. 
In particular, it relates to the usual (symmetric) derivative of $\mathcal{F}_{\mu\nu}$ as
\begin{equation}
n \mathcal{F}_{\mu\nu;\rho}(\pmb{\theta})  =  \partial_{\mu\rho} m(\pmb{\theta})^T C^{-1} \partial_\nu m(\pmb{\theta}) + \partial_{\mu} m(\pmb{\theta})^T C^{-1} \partial_{\nu\rho} m(\pmb{\theta}) \equiv  n\mathcal{F}_{\nu;\rho}^\mu(\pmb{\theta}) + n \mathcal{F}_{\mu;\rho}^\nu(\pmb{\theta}) \ ,
\end{equation}
where the second equivalence holds when all final quantities are evaluated at $\pmb{\theta}=\pmb{\theta_*}$. 
Notice that $\mathcal{F}_{\mu\nu;\rho}$ is fully symmetric under indices permutation while $\mathcal{F}^\mu_{\nu;\rho}$ is solely symmetric under exchange $\nu \leftrightarrow \rho$. 

Inspecting the definition~\eqref{eq:moment}, we wish to calculate perturbatively
\begin{equation}
\int d^N \pmb{\theta} \ f(\pmb{\theta}) \mathcal{P}(\pmb{\theta} | D) \ .
\end{equation}
Here $f$ is an analytic function that can be expanded around $\pmb{\theta_*}$ as
\begin{equation} \label{eq:f}
f(\pmb{\theta}) = f_* + n^{-1/2} f_{;\mu} \delta_\mu + \frac{n^{-1}}{2} f_{;\mu\nu} \delta_\mu  \delta_\nu + \dots \ ,
\end{equation}
where $f_* =  f(\pmb{\theta_*})$. 
Together with~\eqref{eq:p_expand}, we get
{\footnotesize
\begin{align}\label{eq:fp_expand}
f(\pmb{\theta})\mathcal{P}(\pmb{\theta}|\mathcal{F}, \pmb{j}) & = \mathcal{P}(\pmb{\theta_*})  \exp \left( -\frac{1}{2}\delta_\mu \mathcal{F}_{\mu\nu} \delta_\nu + j_\mu \delta_\mu \right) \times  \Bigg\{ f_* +  n^{-1/2} \left[f_{;\mu} \delta_\mu + \frac{f_*}{2} j_{\mu;\nu} \delta_\mu \delta_\nu - \frac{f_*}{4}\mathcal{F}_{\mu\nu;\rho} \delta_\mu \delta_\nu \delta_\rho \right]    \\
& + \frac{n^{-1}}{2}  \left[f_{;\mu\nu} \delta_\mu  \delta_\nu +  \left(\frac{f_*}{3}  j_{\mu;\nu\rho} + f_{;\rho}  j_{\mu;\nu}\right)   \delta_\mu \delta_\nu \delta_\rho - \left(\frac{f_*}{8}\mathcal{F}_{\mu\nu;\rho\sigma} + \frac{f_*}{12} \mathcal{F}_{\nu;\rho\sigma}^\mu  + \frac{f_{;\sigma}}{2}\mathcal{F}_{\mu\nu;\rho}\right)  \delta_\mu \delta_\nu \delta_\rho \delta_\sigma \right] \Bigg\} + \dots \nonumber
\end{align}} 
For definiteness, we start with Lebesgue measure. 
To proceed, it is useful to define a Gaussian generating functional, 
\begin{equation}\label{eq:generating}
G[\pmb{j}] := \int \frac{d^N \pmb{\delta}}{n^{1/2} } \ \exp \left( -\frac{1}{2}\delta_\mu \mathcal{F}_{\mu\nu} \delta_\nu + j_\mu \delta_\mu \right) \propto \exp \left(\frac{1}{2} j_\mu \mathcal{F}^{-1}_{\mu\nu} j_\nu \right) \ .
\end{equation}
Upon integration over $\int d^N\pmb{\delta}/ n^{1/2}$ , we can replace in eq.~\eqref{eq:fp_expand} the exponential by $G$ and the powers of $\delta$'s as $\delta_\mu \rightarrow g_\mu := G^{-1} \, \partial G/\partial{j_\mu} \ , \delta_\mu \delta_\nu \rightarrow g_{\mu\nu} := G^{-1} \, \partial^2 G/(\partial j_\mu \partial j_\nu)$, etc.
For example, the zeroth and unormalised first moments of $\mathcal{P}$ correspond to, respectively, $f=1$ and $f=\pmb{\theta}-\pmb{\theta_*} \equiv \pmb{\tilde \delta}$, yielding
{\footnotesize
\begin{align*}
\mathcal{Z}_\mathcal{P} / G[\pmb{j}] & =  1 +  \frac{n^{-1/2}}{2} \left[ j_{\mu;\nu} g_{\mu\nu} -\mathcal{F}^\mu_{\nu;\rho} g_{\mu\nu\rho}  \right]  + \frac{n^{-1}}{2}  \left[ \frac{1}{3}j_{\mu;\nu\rho}  g_{\mu\nu\rho} - \frac{1}{4}\left(\frac{1}{2}\mathcal{F}_{\mu\nu;\rho\sigma} + \frac{1}{3} \mathcal{F}^\mu_{\nu;\rho\sigma} \right) g_{\mu\nu\rho\sigma} \right]  + \dots \ , \\
\tilde{\mathbb{E}}_{\mathcal{P}}[\tilde \delta_\alpha] /  G[\pmb{j}] & =  n^{-1/2}  g_{\alpha} + \frac{n^{-1}}{2}\left[ j_{\mu;\nu} g_{\alpha\mu\nu} - \frac{1}{2}\mathcal{F}_{\mu\nu;\rho} g_{\alpha\mu\nu\rho} \right]  + \dots 
\end{align*}
}
The normalised 1-moment is then obtained by perturbatively inverting the evidence,
\begin{equation}\label{eq:mean0}
\mathbb{E}_{\mathcal{P}}[\tilde \delta_\alpha] = \frac{\tilde{\mathbb{E}}_{\mathcal{P}}[\tilde \delta_\alpha]}{\mathcal{Z}_\mathcal{P}}  =  n^{-1/2}  g_{\alpha} + \frac{n^{-1}}{2} \bigg[ j_{\mu;\nu}  \left( g_{\alpha\mu\nu} - g_{\alpha} g_{\mu\nu} \right) - \frac{1}{2}\mathcal{F}_{\mu\nu;\rho} \left( g_{\alpha\mu\nu\rho} - g_{\alpha} g_{\mu\nu\rho} \right) \bigg] + \dots 
\end{equation}
The derivatives of $G$ in eq.~\eqref{eq:generating} define the Gaussian moments $g\equiv g[\mathcal{F}, \pmb{j}]$, 
\begin{align}
g_\mu & = \mathcal{F}^{-1}_{\mu\lambda} j_\lambda \ ,
\quad g_{\mu\nu} = g_\mu g_\nu + \mathcal{F}^{-1}_{\mu\nu} \ , \quad
g_{\mu\nu\rho} = g_\mu g_\nu g_\rho 
+ \mathcal{F}^{-1}_{\mu\nu} g_\rho 
+ \mathcal{F}^{-1}_{\mu\rho} g_\nu 
+ \mathcal{F}^{-1}_{\nu\rho} g_\mu \ , \nonumber \\
g_{\mu\nu\rho\sigma} & =\sum_{\text{Sym}(\mu,\nu,\rho,\sigma)} \left( g_{\mu\nu} g_\rho g_\sigma + \mathcal{F}^{-1}_{\mu\nu} g_{\rho\sigma} \right)  \ ,
\end{align}
yielding 
\begin{equation}
g_{\alpha\mu\nu} - g_\alpha g_{\mu\nu} = \mathcal{F}^{-1}_{\alpha\mu} g_\nu + \mathcal{F}^{-1}_{\alpha\nu} g_\mu \ , \quad g_{\alpha\mu\nu\rho} - g_\alpha g_{\mu\nu\rho}
= \sum_{\text{cyc}(\mu,\nu,\rho)}  \mathcal{F}^{-1}_{\alpha\mu} g_{\nu\rho}  \ .
\end{equation}
In appendix~\ref{app:n-2}, we show analogous expressions for the zeroth, first, second, and third moments up to $\mathcal{O}\left(n^{-2}\right)$. 


\section{Efficient parameter estimation}\label{sec:efficient}

In this section, after understanding the origin of the leading-order average bias in the standard definitions for the mean and the mode of marginal posterior distributions, we present unbiased estimators for the mean and the mode. 
In particular, we define an estimator to be efficient if it is unbiased on average at leading order (up to $\mathcal{O}(n^{-2}$)) and with minimal variance. 
We stress that here and in the rest of the paper, `average' refers to the average over data samples as defined by eq.~\eqref{eq:average}, whereas `mean' refers to the mean ($1$-moment) of a posterior distribution of some parameters as defined in eq.~\eqref{eq:moment}.  
For example, the average posterior mean bias refers to the bias in the mean of the posterior distribution averaged over data samples. 

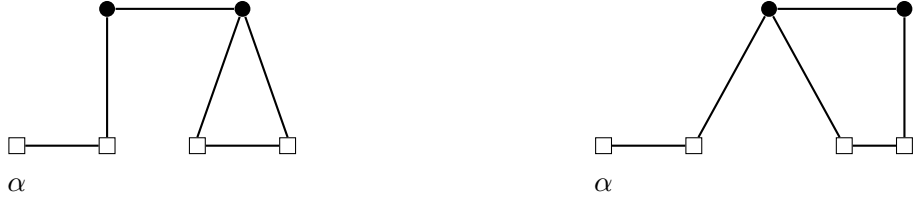
\begin{figure}[h]
\centering

\begin{minipage}{0.48\textwidth}
\centering
\begin{tikzpicture}[xscale=1.0, yscale=1.3]

\node[draw, minimum size=6pt, inner sep=0pt] (sAlpha) at (0,0) {};
\node at (0,-0.4) {\( \alpha \)};

\node[draw, minimum size=6pt, inner sep=0pt] (sMu) at (1.2,0) {};
\draw[thick] (sAlpha) -- (sMu);

\node[fill, circle, minimum size=6pt, inner sep=0pt] (c1) at (1.2,1.4) {};
\draw[thick] (sMu) -- (c1);

\node[fill, circle, minimum size=6pt, inner sep=0pt] (c2) at (3.0,1.4) {};
\draw[thick] (c1) -- (c2);

\node[draw, minimum size=6pt, inner sep=0pt] (sNu) at (2.4,0) {};

\node[draw, minimum size=6pt, inner sep=0pt] (sRho) at (3.6,0) {};

\draw[thick] (c2) -- (sNu.north);
\draw[thick] (c2) -- (sRho.north);

\draw[thick] (sNu) -- (sRho);

\end{tikzpicture}
\end{minipage}
\begin{minipage}{0.48\textwidth}
\centering
\begin{tikzpicture}[xscale=1.0, yscale=1.3]

\node[draw, minimum size=6pt, inner sep=0pt] (sAlpha) at (0,0) {};
\node at (0,-0.4) {\( \alpha \)};

\node[draw, minimum size=6pt, inner sep=0pt] (sMu) at (1.2,0) {};
\draw[thick] (sAlpha) -- (sMu);

\node[draw, minimum size=6pt, inner sep=0pt] (sNu) at (3.2,0) {};

\node[fill, circle, minimum size=6pt, inner sep=0pt] (c1) at (2.2,1.4) {};
\draw[thick] (sMu) -- (c1);
\draw[thick] (sNu) -- (c1);

\node[fill, circle, minimum size=6pt, inner sep=0pt] (c2) at (4.0,1.4) {};
\draw[thick] (c1) -- (c2);

\node[draw, minimum size=6pt, inner sep=0pt] (sRho) at (4.0,0) {};
\draw[thick] (c2) -- (sRho.north);

\draw[thick] (sNu) -- (sRho);

\end{tikzpicture}
\end{minipage}
\caption{Diagrammatic representation of the average mean bias 
$\mathcal{F}^{-1}_{\alpha\mu} \mathcal{F}^{-1}_{\nu\rho} \mathcal{F}_{\mu\nu;\rho} = \mathcal{F}^{-1}_{\alpha\mu} \mathcal{F}^\mu_{\nu;\rho} \mathcal{F}^{-1}_{\nu\rho} + \mathcal{F}^{-1}_{\alpha\mu} \mathcal{F}^\nu_{\mu;\rho} \mathcal{F}^{-1}_{\nu\rho}$. 
See app.~\ref{sec:feyman} for details on the diagram rules. }
\label{fig:feyman_meanbias}
\end{figure}

\subsection{The mean}\label{sec:mean}
Based on the Laplace expansion above, we are now ready to assess the efficiency of the mean ($1$-moment) estimator.  
Namely, we want to quantify potential bias and how closely the variance in the estimator approaches the Cram\'er-Rao bound (or equivalently, differs from the inverse Fisher information). 
The bias in the mean is dominated by the noise/data-dependent term $n^{-1/2} \mathcal{F}_{\mu\nu}^{-1} j_\nu$, keeping in mind that $j_\nu \equiv j_\nu(\Delta_*)$, where $\Delta_* \equiv y-m_*$. 
The efficiency of the moment estimators is however assessed in an average sense, \emph{i.e.}, by averaging over data samples. 
For now let us assume $\braket{\Delta_*} = 0$.\footnote{We will relax this assumption in sec.~\ref{sec:mode}.} 
Hence, the term in $\mathcal{O}(n^{-1/2})$ vanishes. 
Similarly, terms in $\mathcal{O}(n^{-3/2})$ average to $0$ since they are odd in powers of $\Delta$. 
In contrast, since $\braket{\Delta_* \Delta_*} = C$, we have
\begin{align}
\braket{j_\mu j_\nu} & = \mathcal{F}_{\mu\nu} \ , \quad \braket{j_{\rho} j_{\mu;\nu}} = \mathcal{F}^\rho_{\mu;\nu} \ , \label{eq:jj} \\
\braket{g_\mu g_\nu} & = \mathcal{F}^{-1}_{\mu\nu} \ ,  \quad  \label{eq:gg} 
\end{align}
and so on. 
Under sample averaging, the posterior mean estimator with Lebesgue measure, eq.~\eqref{eq:mean0}, is then biased at $\mathcal{O}(n^{-1})$, 
\begin{equation}\label{eq:meanbias}
\braket{\mathbb{E}_{\mathcal{P}}[\tilde \delta_\alpha]}   = - n^{-1} \mathcal{F}^{-1}_{\alpha\mu} \mathcal{F}^{-1}_{\nu\rho} \mathcal{F}_{\mu\nu;\rho} + \mathcal{O}\left(n^{-2}\right) \ . 
\end{equation}
We check our results using diagrammatic representation laid out in app.~\ref{sec:feyman}, leading to the diagrams shown in fig.~\ref{fig:feyman_meanbias}. 
From eq.~\eqref{eq:mean0} and using eq.~\eqref{eq:gg}, we see that the variance of the mean estimator is
\begin{equation}
\braket{\mathbb{E}_{\mathcal{P}}[\tilde \delta_\alpha]\mathbb{E}_{\mathcal{P}}[\tilde \delta_\beta]} = n^{-1} \mathcal{F}_{\alpha\beta}^{-1} + \mathcal{O}(n^{-2}) \ . 
\end{equation}
To sum up, on average at $\mathcal{O}(n^{-1})$, the mean ($1$-moment) estimator is biased but with minimal variance. 
In section~\ref{sec:estimator}, we will consider another definition for the mean estimator, eq.~\eqref{eq:moment}, making use of non-Euclidian measure, such that the bias will start at $\mathcal{O}(n^{-2})$.

\subsection{The mode}\label{sec:mode}
Our findings in sec.~\ref{sec:mean} on the efficiency of the mean estimator rely on the fact that the posterior mode $\pmb{\theta_*}$ coincides with the truth $\pmb{\theta_\dagger}$. 
This was implicitly assumed when stating that $\braket{\Delta_*} = 0$, as $\braket{\Delta_*} = \braket{y}-m_*$. \
In reality, as we show below, the mode itself is biased with respect to the truth~\cite{McCullagh}. 
Because, $\pmb{\theta_*} \neq \pmb{\theta_\dagger}$, we then have $\braket{\Delta_*} = \braket{y}-m_* = m_\dagger - m_* \neq 0$, which in turn has implication for the mean bias, that we elucidate in the next section. 

As previously, we recall that we work in the limit of no model mispecification, flat prior, and consider an asymptotic normal distribution. 
The posterior mode corresponds in this case to the MLE, eq.~\eqref{eq:mle}. 
In the case of \emph{noisy} data samples, averaging over many leads to a non-vanishing $\mathcal{O}(n^{-1})$ bias in the average MLE, that stems from the noise variance.

By definition, the score function at the posterior mode $\pmb{\theta_*}$ vanishes, 
\begin{equation}\label{eq:score}
\partial_\mu \log \mathcal{P}(\pmb \theta | y) \big|_{\pmb{\theta} = \pmb{\theta_*}}  = 0 \ .
\end{equation}
Let $n^{-1/2} \pmb{ \delta_\dagger} = \pmb{\theta_*} - \pmb{\theta_\dagger}$ be the shift of the mode to the ground truth. 
Taylor-expanding for $n^{-1/2} \pmb{ \delta_\dagger}$, assuming it is a small perturbations, eq.~\eqref{eq:score} becomes
\begin{equation}
- \mathcal{F}_{\mu\alpha}\delta^\dagger_\alpha + j_\mu + n^{-1/2}\left[j_{\mu ; \nu}\delta^\dagger_\nu - \frac{1}{2}(\mathcal{F}^{\mu}_{\nu; \rho} + \mathcal{F}^{\nu}_{\rho; \mu} + \mathcal{F}^{\rho}_{\mu; \nu} )\delta^\dagger_\nu\delta^\dagger_\rho \right] + \mathcal{O}(n^{-1}) = 0 \ ,
\end{equation}
where now, instead, all quantities (but $\pmb{\delta_\dagger})$ are evaluated at $\pmb{\theta_\dagger}$. 
At zeroth order in powers of $n^{-1/2}$, we find that $\delta_\alpha^{(0)} = j_\mu \mathcal{F}_{\mu\alpha}^{-1} $. 
Plugging this solution into the $\mathcal{O}(n^{-1/2})$-term then yields
\begin{equation}
\delta^\dagger_\alpha = j_\mu \mathcal{F}^{-1}_{\mu\alpha} + n^{-1/2}\left[ j_{\mu;\nu}j_\rho \mathcal{F}^{-1}_{\rho\nu}\mathcal{F}^{-1}_{\mu\alpha} - \frac{1}{2} (\mathcal{F}^{\mu}_{\nu; \rho} + \mathcal{F}^{\nu}_{\rho; \mu} + \mathcal{F}^{\rho}_{\mu; \nu} ) j_\sigma j_\eta \mathcal{F}^{-1}_{\sigma \nu}\mathcal{F}^{-1}_{\eta \rho}\mathcal{F}_{\mu\alpha}^{-1} \right] + \mathcal{O}(n^{-1}) \ .  
\end{equation}
Defining $\Delta_\dagger = y - m_\dagger$, we have $\braket{\Delta_\dagger} = 0$ and $\braket{\Delta_\dagger \Delta_\dagger} = C$, truly. 
Averaging over data samples and using~\eqref{eq:jj}, we find that the average mode presents a bias at $\mathcal{O}(n^{-1})$, 
\begin{equation}\label{eq:modebias}
\braket{\tilde \delta^\dagger_\alpha} \equiv  n^{-1/2} \braket{\delta_\alpha^\dagger} = -\frac{n^{-1}}{2}\mathcal{F}^{-1}_{\alpha\mu} \mathcal{F}^{-1}_{\nu\rho}\mathcal{F}^\mu_{\nu;\rho} + \mathcal{O}(n^{-2}) \ ,
\end{equation}
where the next-order bias is $\mathcal{O}(n^{-2})$ as terms in $\mathcal{O}(n^{-3/2})$ are odd in powers of $\Delta$.

\subsection{The mean, revisited}\label{sec:mean_revisited}
On average and relative to the truth, as the mode is biased, the mean inherits a new bias, 
\begin{equation}\label{eq:modebias2mean}
\braket{\mathbb{E}_{\mathcal{P}}[\theta_\alpha - \theta^\dagger_\alpha]} = \braket{\mathbb{E}_{\mathcal{P}}[\theta_\alpha - \theta^*_\alpha]} + \braket{\theta^*_\alpha-\theta^\dagger_\alpha} = \braket{\mathbb{E}_{\mathcal{P}}[\tilde \delta_\alpha]} + \braket{\tilde \delta^\dagger_\alpha} \ ,
\end{equation}
where $\braket{\mathbb{E}_{\mathcal{P}}[\tilde \delta_\alpha]}$ is given by eq.~\eqref{eq:meanbias} and $\braket{\tilde \delta^\dagger_\alpha}$ is given by eq.~\eqref{eq:modebias}. 
Besides, the average mode bias introduces further contributions in $\braket{\mathbb{E}_{\mathcal{P}}[\tilde \delta_\alpha]}$ as now $\braket{y} \neq m_*$. 
We show that, on average at $\mathcal{O}(n^{-1})$, a new contribution in $\braket{\mathbb{E}_{\mathcal{P}}[\tilde \delta_\alpha]}$, assuming instead $\braket{y} = m_\dagger$, exactly cancels the additional bias from the mode in eq.~\eqref{eq:modebias2mean}. 
First, we realise that 
\begin{equation}
\Delta_* = y-m_* = y - m_\dagger + m_\dagger - m_* = \Delta_\dagger - n^{-1/2} \, \partial_\nu m \, \delta_\nu^\dagger + \dots \ ,
\end{equation} 
where we have defined $\Delta_\dagger = y-m_\dagger$ and expanded $m_\dagger$ around $\pmb{\theta_*}$. 
Anticipating that terms proportional to odd powers of $\Delta_\dagger$ now truly average to $0$, we shall focus on contributions that arise from the second term in the right-hand side above. 
The source term, eq.~\eqref{eq:source}, receives a correction, 
\begin{equation}
n^{1/2} j_\mu = \partial_\mu m^T  C^{-1}  \Delta_* = n^{1/2}j^\dagger_\mu - \ n^{1/2} \mathcal{F}_{\mu\nu} \ \delta_\nu^\dagger \ , 
\end{equation}
where we used eq.~\eqref{eq:fisher}, the definition of $n\mathcal{F}_{\mu\nu}$, and defined $n^{1/2}j^\dagger_\mu = \partial_\mu m^T  C^{-1}  \Delta_\dagger$. 
From the leading noise term in the mean estimator given in eq.~\eqref{eq:mean0}, we then get the following correction, 
\begin{equation}
\mathbb{E}_{\mathcal{P}}[\tilde \delta_\alpha] \supset n^{-1/2} g_\alpha  = n^{-1/2} \mathcal{F}_{\alpha\mu}^{-1} j_\mu = n^{-1/2} \mathcal{F}_{\alpha\mu}^{-1} j^\dagger_\mu - \ n^{-1/2} \delta_\alpha^\dagger \ . 
\end{equation}
Under sample averaging and using eq.~\eqref{eq:modebias}, we thus find a new bias contribution in the mean estimator at $\mathcal{O}(n^{-1})$, 
\begin{equation}\label{eq:newbias}
\braket{\mathbb{E}_{\mathcal{P}}[\tilde \delta_\alpha]} \supset - \braket{\tilde \delta^\dagger_\alpha} \ .
\end{equation}
To sum up, the total average mean bias \emph{relative to the truth} is the sum of the average mean bias relative to the mode, the average mode bias relative to the truth, eq.~\eqref{eq:modebias2mean}, and the additional correction given in eq.~\eqref{eq:newbias}. 
The two latter terms are opposite of the same quantity (both evaluated at $\pmb{\theta_\dagger}$) so they cancel each others. 
The average mean bias is then simply given by eq.~\eqref{eq:meanbias}. 

\paragraph{Noiseless limit}
Noiseless synthetic data (but with finite covariance) are often useful to assess the performance of the inference. 
In this case, the mode is unbiased, and it is easy to see from eq.~\eqref{eq:mean0} that the average mean bias is
\begin{equation}\label{eq:meanbias_noiseless}
\braket{\mathbb{E}_{\mathcal{P}}[\tilde \delta_\alpha]} = - n^{-1} \mathcal{F}^{-1}_{\alpha\mu} \mathcal{F}^{-1}_{\nu\rho} ( \mathcal{F}^\nu_{\mu;\rho} + \frac{1}{2}\mathcal{F}^\mu_{\nu;\rho}) + \mathcal{O}\left(n^{-2}\right) \ . 
\end{equation}
Compared to eq.~\eqref{eq:meanbias}, they differs by one half of the term proportional to  $\mathcal{F}^\mu_{\nu;\rho}$. 

\paragraph{Gaussian prior}
In presence of an additional Gaussian prior distribution centred on $\pmb{\hat \theta}$ with covariance $\mathcal{C}$, the Fisher matrix and the source term defined in~\eqref{eq:fisher}~and~\eqref{eq:source} are modified to
\begin{align}
n \mathcal{F}_{\mu\nu}(\pmb{\theta}) & \rightarrow \partial_\mu m^T(\pmb{\theta}) C^{-1} \partial_\nu m(\pmb{\theta}) + \mathcal{C}_{\mu\nu}^{-1} \  ,  \label{eq:F_pr}\\
n^{1/2} j_\mu(\pmb{\theta}) & \rightarrow  \partial_\mu m^T(\pmb{\theta}) C^{-1} \Delta_* +  \mathcal{C}_{\mu\nu}^{-1} (\hat \theta_\nu-\theta^*_\nu)  \ . 
\end{align}
The rests of the formulae in sec.~\ref{sec:laplace} are unchanged as additional contributions from the prior are killed when taking derivatives of the Fisher or the source term. 
In contrast, the variance of $j_\mu$ now receives an additional non-vanishing contribution when the prior is not centered on the mode,  
\begin{equation}\label{eq:jj_pr}
\braket{j_\mu j_\nu} = \mathcal{F}_{\mu\nu} + \mathcal{C}_{\mu\rho}^{-1} (\hat \theta_\rho-\theta^*_\rho) \mathcal{C}_{\nu\sigma}^{-1} (\hat \theta_\sigma-\theta^*_\sigma) \ .  
\end{equation}
In practical situations, there are two relevant limits. 
Either the prior is weak compared to the data likelihood, as \textit{e.g.}, when informed by general order-of-magnitude estimates stemming from naturalness or perturbativity of the theory. 
In this case, the prior central value is different to the mode only so within a weak prior variance, therefore leading to negligible shift in the posterior distributions. 
Either, the prior is strong, as \textit{e.g.}, when informed from other experiments constraining a particular subset of the model parameters. 
In sec.~\ref{sec:lss}, we will see practical examples where these two situations appear. 
Therefore, in these two limits, the difference between the prior central value and the mode is generally small, and even more so when squared as appearing in eq.~\eqref{eq:jj_pr}. 
Neglecting this additional contribution, the average mean bias is given by eq.~\eqref{eq:meanbias} but with the replacement~\eqref{eq:F_pr}.

\paragraph{Model mispecification}
When the model is mispecified, \textit{i.e.}, $\braket{\Delta^\dagger} \equiv \delta m \neq 0$, we can estimate the resulting bias in the mean of the distribution. 
At leading order, 
\begin{equation}\label{eq:biastheo}
\braket{\mathbb{E}_{\mathcal{P}}[\delta_\alpha]} = n^{-1/2} \mathcal{F}_{\alpha\mu}^{-1} \ \partial_\mu m^T C^{-1} \delta m + \dots
\end{equation}
This `theoretical' bias is independent of the `statistical' bias encountered above. 
If the size of the theory error $\delta m$ is known, one can then estimate the shift in the mean for the parameters of interest --- even better, rotate to the basis where the Fisher is diagonal such that theoretical bias on principal components can be kept under controlled.\footnote{In the galaxy clustering example given in sec.~\ref{sec:EFTofLSS}, $\delta m$ would be of the size of the two-loop contributions in the EFTofLSS, and one of the principal component would be closely related to $\sim f\sigma_8$. }
Another quantity that could be computed without relying on numerical sampling or being affected by volume projection effects is the Figure of Bias $\mathcal{B}$, defined as $\mathcal{B}^2 = b_\mu \mathcal{F}_{\mu\nu} b_\nu$, 
where $b_\alpha$ is the bias vector given by eq.~\eqref{eq:biastheo}. 
We leave such investigation to future work. \\

In summary, given a posterior distribution, a practical and effective strategy for robust parameter estimation involves both knowledge of the mode and the mean. 
One first finds the posterior mode, and then, uses eq.~\eqref{eq:meanbias}, evaluating its quantities at the mode, to estimate the average shift of the mean relative to the truth. 
In the next section, we present another way of achieving the same results by redefining the volume measure in expectation values. 

\subsection{Non-flat volume measure}\label{sec:estimator}

We show that for an asymptotic normal distribution $\mathcal{P}(\theta|y)$, the mean ($1$-moment) estimator $\mathbb{E}_{\mathcal{P}, \mathcal{M}_\mathcal{H}}[\theta_{\alpha}]$, eq.~\eqref{eq:moment}, with measure
\begin{equation}\label{eq:H}
\mathcal{M}_\mathcal{H}(\pmb{\theta}) \equiv \sqrt{\det \mathcal{H}(\pmb{\theta})} d^N \pmb{\theta} \ , \quad
\mathcal{H}_{\mu\nu}(\pmb{\theta}) := -\partial_{\mu\nu} \log \mathcal{P} ( \pmb{\theta}|y=m_* )  \ ,
\end{equation}
where $m_* \equiv m(\pmb{\theta_*})$, is unbiased and with minimum variance up to $\mathcal{O}(n^{-2})$ under sample averaging. 
Moreover, $\mathcal{M}_\mathcal{H}$ is shown to be invariant under reparametrisation. 
Given such properties, $\mathcal{H}_{\mu\nu}$ can be intuitively thought as the metric tensor of a (pseudo-)Riemannian manifold $\mathcal{\log P}_*$, with $\mathcal{P}_* \equiv \mathcal{P}( \pmb{\theta}|y=m_* )$, living in the parameter space $\pi(\pmb{\theta}) \subset \mathbb{R}^N$.  
Accordingly, $\mathcal{M}_\mathcal{H}$ is then the volume form of $\mathcal{\log P}_*$.\footnote{Following this geometrical interpretation, these statistics are sometimes dubbed `non-flat, `non-Euclidian', or `curved', in opposition to standard `flat' estimators with Lebesgue measure. 
See \textit{e.g.}, ref.~\cite{Kass0} for applications of differential geometry in statistical inference. }
 
\paragraph{Unbiased mean}
Expanding the volume element $\sqrt{\det \mathcal{H}}$ around the mode $\pmb{\theta_*}$, 
\begin{equation}\label{eq:Hexpand}
\frac{1}{2}\log \det \mathcal{H}(\pmb{\theta})  = \frac{1}{2}\log \det \mathcal{H} (\pmb{\theta_*}) + \frac{1}{2}\partial_\mu \log  \det \mathcal{H}(\pmb{\theta})  \Big|_{\pmb{\theta} = \pmb{\theta_*}} (\theta_\mu -\theta_\mu^*) +  \dots \ ,
\end{equation}
leads to new contributions in the expansion of the $p$-moments that can be cast using eq.~\eqref{eq:f}. 
From there, it is easy to see that the $1$-moment defined with measure $\mathcal{M}_\mathcal{H}$, compared to the standard one with Lebesgue measure, eq.~\eqref{eq:mean0}, receives a new contribution at $\mathcal{O}(n^{-1})$,
\begin{equation}\label{eq:meanH}
\mathbb{E}_{\mathcal{P},\mathcal{M}_\mathcal{H}}[\tilde \delta_\alpha] \supset n^{-1}  (g_{\alpha\mu}-g_\alpha g_\mu) \, \frac{1}{2}\partial_\mu \log  \det \mathcal{H}(\pmb{\theta})  \Big|_{\pmb{\theta} =\pmb{\theta_*}} .
\end{equation}
Under sample averaging, we get $\braket{g_{\alpha\mu}-g_\alpha g_\mu} = \mathcal{F}_{\alpha\mu}^{-1}$. 
Moreover, we have 
\begin{equation}\label{eq:Hlinear}
\frac{1}{2}\partial_\mu \log  \det\mathcal{H} \Big|_{\pmb{\theta} = \pmb{\theta_*}} = \frac{1}{2} \mathcal{H}^{-1}_{\nu\rho} \partial_\mu \mathcal{H}_{\nu\rho} \Big|_{\pmb{\theta} =\pmb{\theta_*}} = \mathcal{F}^{-1}_{\nu\rho} \mathcal{F}_{\mu\nu;\rho} \ .
\end{equation} 
Thus, the non-Euclidian measure contributes on average to the posterior mean as
\begin{equation}\label{eq:Hbias}
\braket{\mathbb{E}_{\mathcal{P},\mathcal{M}_\mathcal{H}}[\tilde \delta_\alpha]} \supset  n^{-1} \mathcal{F}^{-1}_{\alpha\mu} \mathcal{F}^{-1}_{\nu\rho} \mathcal{F}_{\mu\nu;\rho} \equiv n^{-1}\mathcal{F}^{-1}_{\alpha\mu} \mathcal{F}^{-1}_{\nu\rho} ( \mathcal{F}^\nu_{\mu;\rho} + \mathcal{F}^\mu_{\nu;\rho}) \ ,
\end{equation}
such that it exactly cancels the average mean bias from the posterior expansion, eq.~\eqref{eq:meanbias}. 
It follows that $\braket{\mathbb{E}_{\mathcal{P},\mathcal{M}_\mathcal{H}}[\tilde \delta_\alpha]}$ is unbiased up to $\mathcal{O}(n^{-2})$. 

\paragraph{Parametrisation invariance}
Let $\pmb \theta$ and $\pmb \phi$ be two model parametrisations related through the Jacobian $J_{\mu\nu} = \partial \theta_\mu /{\partial \phi_\nu}$. 
Given that the metric $\mathcal{H}_{\mu\nu}$ transforms as
\begin{equation}
\mathcal{H}_{\mu\nu}(\pmb \phi) = J_{\mu\alpha}\mathcal{H}_{\alpha\beta}(\pmb \theta)J_{\beta\nu} \ ,
\end{equation}
the volume form $\mathcal{M}_\mathcal{H}$ is invariant under reparametrisation since 
\begin{equation}\label{eq:invariance}
\sqrt{\det \mathcal{H}(\pmb{\phi})} = \sqrt{\det \mathcal{H}(\pmb{\theta})} \det J \ .
\end{equation}

\paragraph{Jeffreys prior}
Another measure one can consider is
\begin{equation}\label{eq:Jeffreys}
\mathcal{M}_\mathcal{F}(\pmb{\theta}) \equiv \sqrt{\det \mathcal{F}(\pmb{\theta})} d^N \pmb{\theta} \  , \quad \mathcal{F}_{\mu\nu}(\pmb{\theta}) = -\braket{\partial_{\mu\nu} \log \mathcal{P}(\pmb{\theta}|y)}  \ .
\end{equation}
$\mathcal{F}_{\mu\nu}$ is the Fisher matrix given by eq.~\eqref{eq:fisher} (setting $n\equiv 1$). 
Also parametrisation invariant, $\sqrt{\det \mathcal{F}}$ is commonly referred as the Jeffreys prior. 
Formally, it is not a probability distribution (said to be improper, as it is does not integrate to one), but rather a volume form that can be chosen a priori, \textit{i.e.}, a prior volume.  
One can check that the first term in eq.~\eqref{eq:Hbias} corresponds to the contribution of the Jeffreys prior to the mean at $\mathcal{O}(n^{-1})$.\footnote{
Since $-\mathcal{F}$ is the average Hessian over sampled log-posteriors, it is tempting to interpret $-\mathcal{H}$ as the Hessian over the average log-posterior. 
However, since $\braket{\log \mathcal{P}(\pmb{\theta}|y)} = \log \mathcal{P}(\pmb\theta|m_\dagger)$, 
we have
\begin{equation}
\mathcal{H}_{\nu\rho}(\pmb{\theta}) = - \partial_{\nu\rho}\log \mathcal{P}(\pmb\theta|m_*) + \partial_{\nu\rho} m(\pmb{\theta})^T C^{-1} (m_* - m_\dagger) \simeq - \partial_{\nu\rho}\log \mathcal{P}(\pmb\theta|m_*)  + n  \mathcal{F}^\mu_{\nu;\rho}(\pmb{\theta}) \tilde \delta^\dagger_\mu  \ , 
\end{equation}
where in the last equality we have expanded $m_\dagger$ around the mode and used the replacements given by eq.~\eqref{eq:nosymfish} and eq.~\eqref{eq:modebias}. 
Since the shift of the mode to the truth $\tilde \delta^\dagger_\mu$ is on average $\mathcal{O}(n^{-1})$, this difference can not be neglected. 
Therefore, $\mathcal{H}_{\nu\rho} \neq -\partial_{\nu\rho}\braket{\log \mathcal{P}(\pmb{\theta}|y)}$. 
} 
In passing, we note that for noiseless synthetic data (but with finite covariance), the relevant measure is $\mathcal{M}_{\mathcal{FH}}(\pmb{\theta}) \equiv \det (\mathcal{F}(\pmb{\theta}) \mathcal{H}(\pmb{\theta}))^{1/4} d^N \pmb{\theta}$, canceling the average mean bias~\eqref{eq:meanbias_noiseless}.

\paragraph{Practicalities}
In summary, we have two possibilities to robustly estimate model parameters given a posterior distribution. 
For both, we first require knowledge of the posterior mode. 
Following one possible route, one computes the shift~\eqref{eq:meanbias} of the mean to the true values, where quantities are evaluated on the mode. 
This procedure then simply appears as a post-debiasing correction, and we will refer to it as so from here on. 
Second possibility, one computes the measure~\eqref{eq:H}, that depends explicitly on the mode, for each posterior samples.  
For cases where some nuisance parameters are linear, one can use properties of Gaussian likelihoods to analytically marginalise over them, reducing the computational complexity of numerical sampling. 
This is detailed in app.~\ref{app:marg}. 
In the following, we will check experimentally on examples in LSS analysis the agreement between the two procedures.

\paragraph{Finale: the Mode, the Mean, and the Ugly}
The mean alone can't find the truth without the help of the `ugly' measure, which relies on the mode.
But can the mean and the `ugly' do without the mode?\footnote{As the story goes, only the Good knows where to dig...}
This question may appeal to some self-proclaimed Bayesians who insist that no prior assumptions should depend on the data from which we seek to extract information.
To such concerns, we stress that our starting point is the posterior distribution, assumed to be given to us, that has been estimated for some model parameters describing the observations.
The relevant question is, from this posterior, how do we build credence in our theory?

For that, we, in any case, must rely on the \emph{data} living on the posterior manifold. 
Expectation values are built from the posterior \emph{data}, which inherently embeds the mode, the Fisher, and so on. 
Since the parameter estimation we advocate depends solely on the posterior \emph{data}, it is not a contradiction to make use of the mode or the Fisher when deriving marginalised statistics. 
Crucially, we are not updating the prior based on posterior \emph{data}. Rather, we are adopting a prescription for projecting the posterior living in a high-dimensional parameter space onto the physical dimensions of interest.

To be precise: the integration measure is not a prior, nor a distribution. 
It is simply a choice to be made when defining expectation values. 
As such, the integration measure only affects derived quantities such as marginal posteriors, credible intervals, and mean values — that is, the quantities we quote when reporting physical results.
The posterior distribution itself remains unchanged by the choice of how we decide to present our results. 
Indeed, in practice, when combining two experiments, we sample again the new posterior formed by combining their likelihoods --- not by combining their marginal posteriors.

On the other hand, it is true that the Jeffreys measure~\eqref{eq:Jeffreys}, as constructed from the Fisher, can be defined irrespective to the observed data (albeit requiring knowledge of the covariance): it does not depends on the mode. 
In that sense, it stands as a true prior volume: a volume form that can be specified prior to any inference. 
May this final configuration appeal to some, in the following we argue that in some cases, the Jeffreys prior measure mitigates most of the volume projection effects. 

\subsection{Large-$N$ enhancement}\label{sec:Nbias}

To understand quantitatively what is controlling the size of the average mean bias, let us go back to our toy model discussed in sec.~\ref{sec:toy}, with its generalisation given at eq.~\eqref{eq:toy2}. 
Defining $u_\mu \equiv \partial_\mu m$ and the coupling constants $\alpha_{\mu\nu} \equiv \partial_{\mu\nu}m$, the theory model reads
\begin{equation}\label{eq:toy3}
m(\pmb{\theta}) = m_* + u_\mu \delta_\mu + \alpha_{\mu\nu}\delta_\mu\delta_\nu + \dots \ ,
\end{equation}
where all quantities (but $\delta$'s) are evaluated on the mode. 
The Fisher and its (non-symmetric) derivative then read
$\mathcal{F}_{\mu\nu} = u_\mu u_\nu$  and $\mathcal{F}^\mu_{\nu;\rho} = u_\mu \alpha_{\nu\rho}$, respectively, where we have set $C^{-1} \equiv 1$ without loss of generality. 
We can consider that we work in the basis where the Fisher is diagonal, $\mathcal{F}_{\mu\nu} \equiv \delta_{\mu\nu}^K$, where $\delta_{\mu\nu}^K$ is the Kronecker delta function, such that $U = \{ u_\mu \}_{\mu=1}^N$ forms a basis of orthonormal vectors. 
The average posterior mean bias is then given by 
\begin{equation}\label{eq:largeN}
\braket{\mathbb{E}_{\mathcal{P}}[\tilde \delta_a]} = - \mathcal{F}^{-1}_{a\mu} \mathcal{F}^{-1}_{\nu\rho} ( \mathcal{F}^\nu_{\mu;\rho} + \mathcal{F}^\mu_{\nu;\rho}) = \delta_{a\mu}^K \delta_{\nu\rho}^K (u_\nu \alpha_{\mu\rho} + u_\mu \alpha_{\nu\rho}) = u_\nu \alpha_{a\nu} + u_{a} \alpha_{\nu\nu} \ .
\end{equation}
Note that the average mean bias involves a sum over all $N$ parameters. 
This implies that, although the leading $n^{-1/2}$-noise term in eq.~\eqref{eq:mean0} remains order one in units of standard deviation $\sigma$, the $n^{-1}$-bias term can be larger than the noise term without contradiction. 
Taking all coupling constants to be order $\alpha$, the expansion parameter $\epsilon \sim \alpha/\sigma \ll 1$ has to remain small to ensure asymptotic convergence. 
Meanwhile, the relative mean bias scales as $\sim N\alpha/\sigma \sim N \epsilon$, and thus become parametrically large as $N \gg 1$.

\paragraph{Linear nuisance parameters}
In many situations in physics, it is sufficient to consider nuisance parameters entering at most linearly in the modeling, meaning their self-couplings $\alpha_{\nu\nu}$ are zero (see \textit{e.g.,} the examples given in the section below). 
Their coupling to the parameters of interest $\alpha_{\mu\nu} \delta_\mu \delta_\nu$, however, may not be parametrically suppressed relative to the signal $u_{\nu} \delta_\nu$, where $\mu$ denotes a parameter of interest and $\nu$ ($\neq \mu$) a nuisance parameter.  
Said differently, this corresponds to the situation where the traceless part of $\alpha_{\mu\nu}$ dominates over its trace. 
When the number of nuisances $N'$ is large compared to the number of parameters of interest, and thus comparable to the total number of parameters ($N' \sim N$), 
the first term on the r.h.s. of eq.~\eqref{eq:largeN} is enhanced relative to the second for large $N$. 
This enhanced bias corresponds to the projection of all couplings $\alpha_{a\nu}$ onto the Fisher eigenvectors $u_\nu$. 
One can check that this contribution is cancelled by the Jeffreys prior measure. 
In such cases, where the second term on the r.h.s. of eq.~\eqref{eq:largeN} can be neglected, the average mode bias is likewise expected to remain small. 


\section{An \'etude on galaxies at long distances}\label{sec:lss}

Recently, prior volume projection effects have attracted a lot of attentions in cosmology, and especially within full-shape analyses of galaxy clustering data~\cite{Ivanov:2019pdj,DAmico:2022osl,Carrilho:2022mon,Simon:2022lde,Donald-McCann:2023kpx,Zhao:2023ebp,Holm:2023laa,Maus:2023rtr,Maus:2024dzi,Zhang:2024thl,DESI:2024jis,Paradiso:2024yqh}.\footnote{See also discussions in related topics: cosmic microwave background~\cite{Planck:2013nga}, weak lensing~\cite{Handley:2019wlz,Joachimi:2020abi,DES:2021rex,Hadzhiyska:2023wae}, beyond $\Lambda$CDM extensions~\cite{Gariazzo:2018pei,Diacoumis:2018ezi,Herold:2021ksg,Cruz:2023cxy,Holm:2022kkd,Camarena:2023cku,Chebat:2025kes}, or broader contexts in astrophysics~\cite{Robnik:2022amb,Handley:2019wlz,Bayer:2020pva,Garcia-Bellido:2023yqk}.}
Because galaxy formation is complex, the most general way to describe the cosmological large-scale structure relies on a perturbative approach: the Effective-Field Theory of Large-Scale Structure (EFTofLSS)~\cite{Baumann:2010tm,Carrasco:2012cv,Senatore:2014eva}. 
In the EFTofLSS, the complicated nonlinear short-scale physics is enclosed in a set of Wilsonian coefficients upon integrating out the short scales that are beyond near-linear scales. 
The latter correspond to the perturbative regime where the galaxy density and velocity fields are $\lesssim \mathcal{O}(1)$. 
Because at sufficiently long distances the only symmetry at hand is the equivalence principle~\cite{Weinberg:2003sw,Peloso:2013zw,Kehagias:2013yd,Creminelli:2013mca}, at each order in perturbations comes a significant number, although finite, of free functions parametrising all the possible gravitational feedbacks between long and short scales. 
For example, in the lowest-order Fourier $N$-point functions, namely the power spectrum and bispectrum, in redshift space, at third order in perturbations there are a total of $10$ Wilsonian coefficients~\cite{Perko:2016puo} while at fourth order there are $41$~\cite{DAmico:2022ukl}. 
These `nuisance' parameters have to be properly marginalise over in the fit to the galaxy clustering data together with the cosmological parameters of interest. 
As the nuisances are numerously coupled to the cosmological parameters, their marginalisation can lead to sizeable volume projection effects~\cite{DAmico:2022osl,Simon:2022lde}. 
Let us elaborate on this. 

\paragraph{Motivationale}
Massive objects in the sky such as galaxies form a map of the cosmos in which we see complex, highly nonlinear structures. 
At the largest scales, the Universe enjoys the simplest symmetry: diffeomorphism invariance. 
Building on this observation, an effective field theory (EFT) allows us to organise the expansions of the underlying density and velocity fields of collapse structures at sufficiently long distances. 
Because gravity couples all Fourier modes $k$, the fields are intrinsically non-Gaussian at all scales.  
This suggests that, at finite $\epsilon \sim k/k_{\rm NL}$ in the asymptotic expansion where $k_{\rm NL}^{-1}$ is the nonlinear (renormalisation) scale, \textit{i.e.}, for a given range of scales that we can probe, there is a sense in pushing to higher order in perturbations since this would allow us, in principle, to access more information beyond the two-point correlation. 
This is somewhat a rather peculiar situation. 
Usually, pushing perturbation theory goes in hand with probing shorter distances, \textit{i.e.}, cranking up $\epsilon$. 
This is useful until the point where the data becomes dominated by the noise or worse, when asymptotic convergence is lost. 
In cosmology, maps of the sky actually give us access to the fields themselves. 
This means that, in principle, additional information residing in the tower of $N$-point functions is to be retrieved by cranking up the computed order in perturbation theory while keeping $\epsilon$ fixed --- a good guarantee of convergence. 
In that sense, we are \emph{adding data} by accessing higher $N$-point functions. 
However, we stress that higher-order corrections come at the cost of introducing an increasingly large number of Wilsonian coefficients. 
We are thus in the following situation: an increasingly large prior volume from an increasingly large number of nuisance parameters when projected onto the marginal posteriors of the quantities of interest is leading to increasingly large \emph{statistical biases}, even though the overall data volume increases. 
This situation motivates us in revisiting the construction of efficient statistics to report as faithfully as possible measurements of cosmological parameters.

Although more data is made accessible as we push the perturbation theory, one may speculate that the total signal-to-noise ratio weighted by an increasing theoretical noise will not improve. 
Even if it turns out to be the case, yet it does not mean that what we are doing is meaningless. 
We can answer using the following picture. 
In all $N$-point functions, baryonic acoustic oscillations (BAO), carrying an important part of the cosmological information that we are after, sit on top of a large broadband signal. 
Suppose that most of the BAO signal can be modelled non-perturbatively without extra parameters, which is partly true as most of the smearing of BAO is due to long-wavelength displacements that can be resummed~\cite{Porto:2013qua,Senatore:2014via}. 
Let us forget also a moment the tantalising possibility of reconstructing the linear BAO faithfully from displacing the observed field in a well thought-out manner (see \textit{e.g.},~ref.~\cite{Eisenstein:2006nk,White:2015eaa}). 
To access the full information residing in the BAO, the broadband signal needs to be properly marginalised by pushing the perturbation theory that describe it. 
This example suggests that the cosmological signal we are after cannot be in principle fully degenerate with the one that is described with the help of nuisance parameters. 
Of course in practice there will be a point where adding more accuracy will only bring a marginal gain in precision.  
At the present, however, preliminary results probing the information in LSS beyond the two-point correlation function seem to indicate that we have not quite saturated the bound (see \textit{e.g.}, ref.~\cite{DAmico:2022osl,Nguyen:2024yth,Spezzati:2025zsb}). 
Already at this stage, $\mathcal{O}(1)$-biases in marginal posteriors of the cosmological parameters due to prior volume projection have been observed. 
This is the practical situation that we aim to address in this paper. 
In the following, we will find that the non-flat volume measures that we have identified allow us to recover cosmological parameters faithfully in LSS analyses. 

\subsection{EFTofLSS analyses and projection effects}\label{sec:EFTofLSS}

Our goal in this section is to highlight how volume projection effects arise in marginal posteriors from LSS analyses. 
To this end, we begin with deliberately simplified, unrealistic assumptions that help clarify what is parametrically controlling the size of the resulting bias. In the following section, we will confront these insights to more realistic setups, computing the posteriors numerically with the aid of MCMC sampling. 

\paragraph{One-loop power spectrum}
Schematically, the prediction for the galaxy density field goes like~\cite{McDonald:2009dh,Senatore:2014eva,Angulo:2015eqa,Assassi:2014fva,Mirbabayi:2014zca}
\begin{equation}\label{eq:dg}
\delta_g(\pmb k, t) = \sum_{i=1}^N b_{i}(t) \mathcal{O}^{(n)}_{i}(\pmb k, t) \  ,
\end{equation}
where $b_{i}(t)$ are the Wilsonian coefficients parametrising our ignorance of short-scales physics and $\mathcal{O}^{(n)}_i$ are scalar operators constructed from all possible contractions of Galilean-invariant fields and spatial derivatives, where the order in perturbations $n$ counts as powers of fields and derivatives. 
Here $\pmb k$ is a Fourier mode such that $|\pmb k | < k_{\rm NL}$, where $k_{\rm NL}^{-1}$ is the nonlinear scale above which the EFTofLSS is predictive. 
$N$-point functions are constructed by taking spatial expectation values of $N$ powers of the galaxy fields at various locations on the sky. 
In the EFTofLSS, these are organised into loop expansions. 
We consider in this work the power spectrum, 
\begin{equation}
(2\pi)^3 \delta_D(\pmb k + \pmb{k'}) \, P(k) \equiv \braket{\delta_g(\pmb k) \delta_g (\pmb{k'})} \, ,
\end{equation}
where the Dirac delta distribution $\delta_D$ enforces translation invariance, and we have dropped the time dependence from the notation. 
Writing $\delta_g = \delta_g^{(1)} + \delta_g^{(2)} + \delta_g^{(3)} + \dots$, the power spectrum at one loop reads
\begin{equation}\label{eq:2pt}
\braket{\delta_g(\pmb k) \delta_g (\pmb{k'})} = \braket{\delta_g^{(1)}(\pmb k) \delta_g^{(1)} (\pmb{k'})} + \braket{\delta_g^{(2)}(\pmb k) \delta_g^{(2)} (\pmb{k'})} + 2 \braket{\delta_g^{(3)}(\pmb k) \delta_g^{(1)} (\pmb{k'})} + \dots \ ,
\end{equation}
where we have assumed that the linear matter fluctuations $\delta^{(1)}(\pmb k)$ are Gaussian distributed with variance $P_{\rm lin}(k)$, so that the diagrammatic structure follows Wick theorem for Gaussian fields. 
The first term of the r.h.s. is the linear contribution and the second and third make for the one-loop correction, a functional of $P_{\rm lin}$ and a polynomial in the EFT parameters $b_{i}$'s. 
Neglecting redshift-space distortions in this discussion for simplicity (although they play a crucial role in determining cosmological parameters in realistic analyses as presented below), the galaxy power spectrum is then a function of the norm $k$ of the vector $\pmb{k}$ and reads
\begin{equation}\label{eq:pk}
P(k) = b_1^2 P_{\rm lin}(k) + \sum_{i,j} b_i b_j P^{ij}_{\rm 1loop}(k) \ , 
\end{equation}
where we have explicitly factorised out the dependence of the loop on the EFT parameters $b_{i}$ such that $P^{ij}_{\rm 1loop}(k)$ are $k$-dependent functions that depend only on the cosmological parameters. 

\paragraph{Parameter coupling}
Introducing a rescaling amplitude parameter $A$ such that $P_{\rm lin} \rightarrow A P_{\rm lin}$, we consider the combination $\theta_1 \equiv b_1 A^{1/2}$ as our proxy for the cosmological parameter that we want to measure. 
For simplicity, we consider that $P^{ij}_{\rm 1loop}$ scales also as $A$, such that we can conveniently define $\theta_i \equiv b_i A^{1/2}$ for all $i = 1, \dots, N$. 
Our simplified model for the power spectrum is
\begin{equation}\label{eq:toy4}
P(k) =  \theta_\mu \theta_\nu P^{\mu\nu}(k) \ ,
\end{equation}
where $P^{11} \equiv P_{\rm lin} + P^{11}_{\rm 1loop}$, and $P^{\mu\nu} \equiv P^{\mu\nu}_{\rm 1loop}$ otherwise. 
It is then easy to see that eq.~\eqref{eq:toy4} once expanded around the mode reduces to eq.~\eqref{eq:toy3} with the identifications $u_\mu \equiv \theta_\nu P^{\mu\nu}$ and $\alpha_{\mu\nu} \equiv P^{\mu\nu}$, where all quantities are evaluated on the mode. 

In reality, only two nonlinear EFT parameters (except $b_1$) appear with $\alpha_{\nu\nu} \equiv P^{\nu\nu}_{\rm 1loop} \neq 0$ (through the 22-diagram), while most others appear only linearly but coupled to $b_1$, \textit{i.e.}, $\alpha_{\nu\nu} = 0$ for $\nu \neq 1$ and $\alpha_{1\nu} \neq 0$. 
This suggests that, as discussed in sec.~\ref{sec:Nbias}, the Jeffreys volume measure is good enough to recover faithfully cosmological parameters in EFTofLSS analyses. 
This is indeed what we will find in the realistic analysis setups presented below.

\subsection{Debiasing cosmological inference}\label{sec:potatoes}

In this section, we assess how different choices of volume measures affect marginal posteriors and credible intervals in cosmological inference from noiseless synthetic galaxy clustering data. 

\begin{table*}[!ht]
\centering
\begin{tabular}{
c
S[table-format=1.2]  
S[table-format=1.2]  
S[table-format=1.3]  
S[table-format=1.0]  
S[table-format=4.0]  
S[table-format=2.1]  
}
\toprule
{Bin} &
{$z_{\min}$} &
{$z_{\max}$} &
{$z_{\mathrm{eff}}$} &
{$V_{\mathrm{eff}}\,[\mathrm{Gpc}^{3}]$} &
{Area\,[deg$^{2}$]} &
{$P_{0}\,[10^{3}\,(h^{-1}\,\mathrm{Mpc})^{3}]$} \\
\midrule
1 & 0.20 & 0.43 & 0.32 &  1 &  3000 & 10.0 \\
2 & 0.20 & 0.43 & 0.32 &  2 &  5000 & 10.0 \\
3 & 0.43 & 0.70 & 0.57 &  2 &  3000 & 10.0 \\
4 & 0.43 & 0.70 & 0.57 &  5 &  5000 & 10.0 \\
\bottomrule
\end{tabular}
\caption{\textbf{BOSS mock survey configurations} ---
The survey footprint totals \(\sim\)\,\SI{8000}{\deg\squared} and is split into one, two, or four observational bins.  
The full four–sky set‐up is shown here. 
The two-sky division corresponds to the low-$z$ and high-$z$ bins of BOSS, each of which is further split into north and south galactic cuts in the four-sky case. 
For comparison analyses, \textit{one sky} corresponds to 
Bin 4 but with $V_{\rm eff}=10 \, \textrm{Gpc}^3$ and $8000 \, \textrm{deg}^2$ sky area, and \textit{two skies} corresponds to 
Bin 2 but with $V_{\rm eff}=3 \, \textrm{Gpc}^3$ and Bin 4 but with $V_{\rm eff}=7 \, \textrm{Gpc}^3$, both on a $8000 \, \textrm{deg}^2$ sky. 
}
\label{tab:boss_setup}
\end{table*}

\begin{table*}[!ht]
\centering
\begin{tabular}{
c
S[table-format=1.2]  
S[table-format=1.2]  
S[table-format=1.3]  
S[table-format=2.0]  
S[table-format=5.0]  
S[table-format=1.1]  
S[table-format=2.0]  
}
\toprule
{Bin} &
{$z_{\min}$} &
{$z_{\max}$} &
{$z_{\mathrm{eff}}$} &
{$V_{\mathrm{eff}}\,[\mathrm{Gpc}^{3}]$} &
{Area\,[deg$^{2}$]} &
{$P_{0}\,[10^{3}\,(h^{-1}\,\mathrm{Mpc})^{3}]$} \\
\midrule
1 & 0.10 & 0.40 & 0.295 &  4 & 14000 & 9.2 \\
2 & 0.40 & 0.60 & 0.510 &  8 & 14000 & 8.9  \\
3 & 0.60 & 0.80 & 0.706 & 12 & 14000 & 8.9  \\
4 & 0.80 & 1.10 & 0.930 & 15 & 14000 & 8.4 \\
5 & 0.80 & 1.10 & 0.930 &  8 & 14000 & 8.4 \\
6 & 1.10 & 1.60 & 1.317 & 12 & 14000 & 2.9 \\
7 & 0.80 & 2.10 & 1.491 &  4 & 14000 & 5.0 \\
\bottomrule
\end{tabular}
\caption{\textbf{DESI Year-6 mock survey configurations} ---
The Y6 footprint covers \SI{14000}{\deg\squared} ($\sim 1/3$ of the full sky)
and is divided into \(n_{\rm sky}=7\) tomographic skies.  
For each bin we list the minimum and maximum survey redshift
(\(z_{\min},\,z_{\max}\)), the effective redshift \(z_{\rm eff}\), the comoving
effective volume \(V_{\rm eff}\), the survey area, and the amplitude of the power spectrum \(P_{0}\) used to estimate the shot noise contribution to the covariance. 
These specifications are extrapolated from refs.~\cite{DESI:2024aax,2503.14738}. 
}
\label{tab:desi_y6_setup}
\end{table*}

\paragraph{BOSS and DESI mock synthetic data analyses}
We consider two realistic setups: a Stage 3-like LSS survey, with total effective volume $V_{\rm eff} = 10 \, \textrm{Gpc}^3$, mimicking SDSS/BOSS~\cite{BOSS:2016wmc}, and a Stage 4-like LSS survey, with $V_{\rm eff} = 63 \, \textrm{Gpc}^3$, approximating DESI Year 6~\cite{DESI:2024aax}. 
Survey characteristics, in particular used to compute the covariance (within the standard Gaussian approximation), are summarised in tables~\ref{tab:boss_setup}~and~\ref{tab:desi_y6_setup}. 
Our synthetic data and model consist of a set of the first three even multipoles $(\ell = 0, 2,4)$ of the power spectrum in redshift space predicted from the EFTofLSS at one loop~\cite{Perko:2016puo}, generated using \PyBird\footnote{\url{https://github.com/pierrexyz/pybird}}~\cite{DAmico:2020kxu}, for each sky listed in the tables. 
We restrict the analysis to the range $[k_{\rm min},k_{\rm max}] = [0.01, 0.20]  \, \hinvMpc$ for all three multipoles and neglect observational effects, as these do not qualitatively impact our conclusions.  
The one-loop redshift-space
power spectrum that we consider in this work depends on twelve EFT parameters and is summarised in our companion paper~\cite{paper1}. 
Linear parameters are analytically marginalised as described in app.~\ref{app:marg}, while we explicitly sample over the remaining nuisance parameters, alongside the cosmological parameters of interest. 
For each bin described in tables~\ref{tab:boss_setup}~or~\ref{tab:desi_y6_setup}, we consider one independent set of EFT parameters. 
We impose broad, order-of-magnitude priors on their size, employing Gaussians centred on $0$ with widths of $\sim \mathcal{O}(b_1)$, motivated by naturalness considerations and inspired by refs.~\cite{DAmico:2019fhj,Simon:2022lde,2411.12021}. 
We adopt the same priors listed in table~3 of our companion paper~\cite{paper1}, with two modifications: (\textit{i}) we fix the next-to-next-to-leading order redshift counterterms to zero, rather than varying them, and (\textit{ii}) instead of varying $b_2$ and $b_4$ individually, we vary the linear combination $c_2 = (b_2 + b_4)/\sqrt{2}$ with a Gaussian prior centered at zero and a width of 5, while setting $c_4 \equiv 0$ in accordance with ref.~\cite{DAmico:2019fhj}. 
These choices do not affect qualitatively our conclusions, given that we are dealing with synthetic data. 

For all analyses, we generate synthetic data using the same fiducial cosmological and EFT parameter values: 
$(\omega_{\rm b} = 0.02235, \omega_{\rm cdm} = 0.120, h=0.675, \ln(10^{10}A_s) = 3.044, n_s=0.965)$, no neutrinos, and $(b_{1} = 1.9542, b_{2} = 0.4173, b_{3} = -0.3686, b_{4} = 0.4173, c_{\rm ct} = 0.1843, c_{r,1} = -0.8477, c_{r,2} = -0.8141, c_{e,0} = 0.0, c_{e,1} = 0.0, c_{e,2} = -1.6279, c_{r,4} = 0.0, c_{r,6} = 0.0)$ (see ref.~\cite{paper1} for a description of the EFT parameters). 
The mean number density $\bar n$ is rescaled according to tables~\ref{tab:boss_setup}~and~\ref{tab:desi_y6_setup} for each sky. 
To compute the one-loop redshift-space galaxy power spectrum, we use the \JAX implementation of \PyBird, accelerated via neural networks as developed in our companion paper~\cite{paper1}. 
We also employ a modified \JAX-compatible version of \texttt{Symbolic-Pk}~\cite{2311.15865,2410.14623} for the linear matter power spectrum, enabling highly efficient cosmological inference. 
Crucially, the automatic differentiation of \JAXBird allows rapid and robust computation of marginal posteriors and credible intervals under non-flat measures, which involves, for example, the Fisher matrix. 
We sample the posterior distributions using the ensemble MCMC sampler \texttt{emcee}\footnote{\url{https://emcee.readthedocs.io/}}~\cite{1202.3665}, for which convergence is monitor through integrated autocorrelation time over the chains. 

\paragraph{Volume measure comparison}
For each analysis setup, we compare four volume-measure prescriptions for obtaining marginal posteriors and credible intervals: 
\begin{itemize}
    \item \textit{Flat measure}: the standard choice where the integration measure in the expectation values~\eqref{eq:moment} is Lebesgue. 
    Marginals are inferred directly from MCMC samples assuming flat priors on cosmological parameters. 
    \item \textit{Post-debiasing correction}: flat-measure marginals shifted by the mean bias calculated in sec.~\ref{sec:efficient}. 
    For simplicity, we consider only the correction associated with the Jeffreys measure, \textit{i.e.}, the first term in eq.~\eqref{eq:meanbias_noiseless}. 
    \item \textit{Jeffreys measure $\mathcal{M}_{\mathcal{F}}$}: implemented by adding to each sampled log-posterior points a log-measure weight $\frac{1}{2}\log \mathcal{F}$ as defined in eq.~\eqref{eq:Jeffreys}. 
    \item \textit{Optimal measure $\mathcal{M}_\mathcal{H}$}: alternative mode-reliant measure $\mathcal{M}_\mathcal{H} = |\mathcal{H}|^{1/2}d^N\pmb\theta$ as defined in eq.~\eqref{eq:H}. 
    As we work with noiseless synthetic data, we actually consider the hybrid measure $\mathcal{M}_\mathcal{FH} \equiv (\mathcal{M}_\mathcal{F}\mathcal{M}_\mathcal{H})^{1/2} = (|\mathcal{F}||\mathcal{H}|)^{1/4}d^N\pmb\theta$. 
\end{itemize}
In practice, log-measure weights must be computed at each parameter point sampled. Since $\mathcal{F}_{\mu\nu}$ and $\mathcal{H}_{\mu\nu}$ can be evaluated as Hessians of log-functions, we exploit the auto-differentiability of the likelihoods developed in our companion paper~\cite{paper1}.
For reference, an inference from the DESI-like mock, run with 64 ensemble walkers and scanning 19 parameters, produces a converged chain in approximately 7 minutes on a single NVIDIA A100 Tensor Core GPU under the standard flat measure. Including the additional log-measure weight (for either $\mathcal{M}_\mathcal{F}$  or $\mathcal{M}_\mathcal{FH}$) increases the runtime to around 35 minutes, which remains reasonably quick.

\begin{figure}[!h]
    \centering
    \footnotesize
    \includegraphics[width=0.48\textwidth]{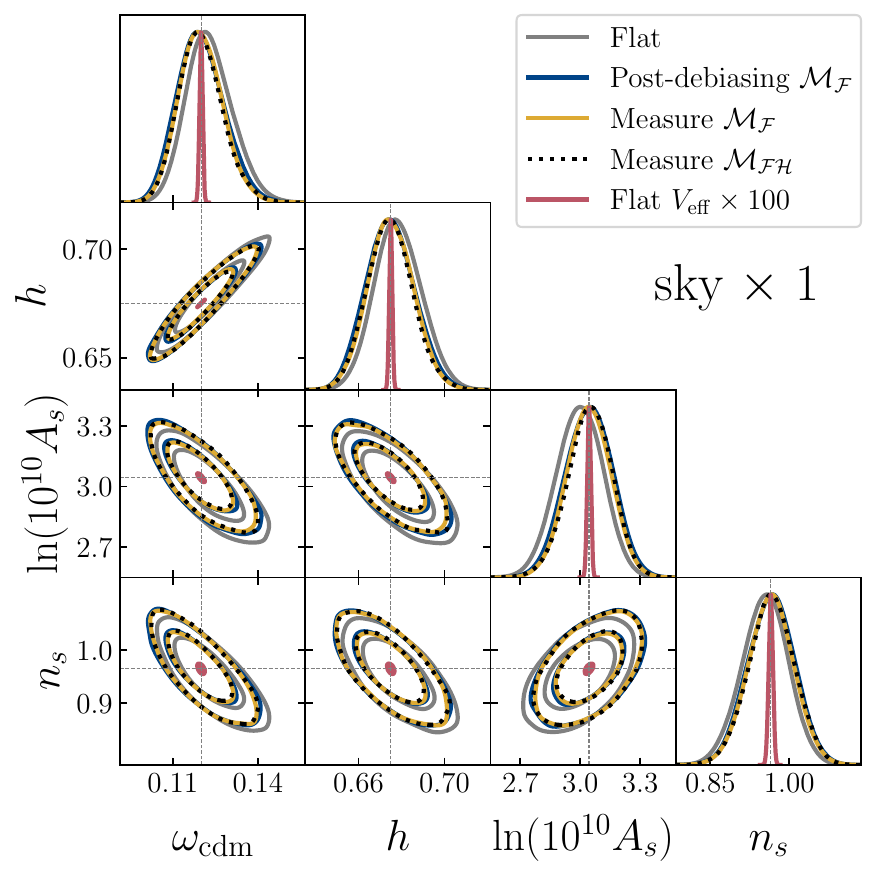}
    \includegraphics[width=0.48\textwidth]{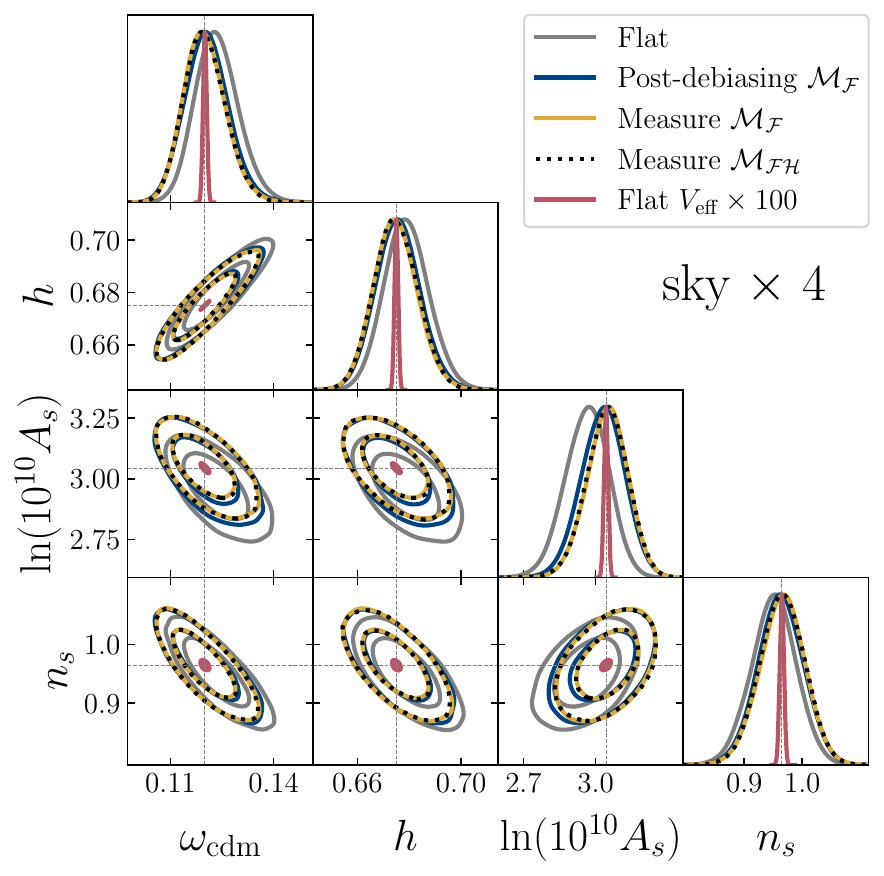}\\
    \vspace{.3cm}
    \begin{tabular}{l l
                S[table-format=1.2]  
                S[table-format=1.2]  
                S[table-format=1.2]  
                S[table-format=1.2]} 
    \toprule
    \textbf{Sky setup} & \textbf{Volume measure}
      & {$\omega_{\mathrm{cdm}}$}
      & {$h$}
      & {$\ln(10^{10}A_s)$}
      & {$n_s$} \\
    \midrule
    \multirow{4}{*}{\textit{One sky}}
      & Flat measure                          &  0.36 &  0.32 & -0.41 & -0.25 \\
      & Post-debiasing $\mathcal{M}_\mathcal{F}$       & -0.00 & -0.01 & -0.00 &  0.01 \\
      & Measure $\mathcal{M}_\mathcal{F}$              & -0.00 &  0.00 & -0.01 &  0.00 \\
      & Measure $\mathcal{M}_{\mathcal{F}\mathcal{H}}$ & -0.01 & -0.03 &  0.04 &  0.03 \\
     \multirow{1}{*}{\textit{$V_{\mathrm{eff}}\times100$}}
      & Flat measure                         &  -0.01 & -0.01 & 0.00 &  -0.01 \\
    \addlinespace
    \multirow{4}{*}{\textit{Two skies}}
      & Flat measure                            &  0.42 &  0.36 & -0.56 & -0.30 \\
      & Post-debiasing $\mathcal{M}_\mathcal{F}$       &  0.05 &  0.05 & -0.06 & -0.01 \\
      & Measure $\mathcal{M}_\mathcal{F}$              & -0.03 & -0.05 &  0.09 &  0.05 \\
      & Measure $\mathcal{M}_{\mathcal{F}\mathcal{H}}$ & -0.02 & -0.03 &  0.05 &  0.02 \\
     \multirow{1}{*}{\textit{$V_{\mathrm{eff}}\times100$}}
      & Flat measure                            & 0.04 &  0.04 & -0.05 & -0.04 \\
    \addlinespace
    \multirow{4}{*}{\textit{Four skies}}
      & Flat measure                       &  0.56 &  0.46 & -0.89 & -0.36 \\
      & Post-debiasing $\mathcal{M}_\mathcal{F}$       &  0.09 &  0.07 & -0.15 & -0.06 \\
      & Measure $\mathcal{M}_\mathcal{F}$              &  0.00 & -0.03 &  0.05 &  0.01 \\
      & Measure $\mathcal{M}_{\mathcal{F}\mathcal{H}}$ & -0.00 & -0.02 &  0.04 &  0.01 \\
     \multirow{1}{*}{\textit{$V_{\mathrm{eff}}\times100$}}
      & Flat measure                       &  0.04 &  0.04 & -0.06 & -0.03 \\
    \bottomrule
    \end{tabular}
    \caption{\textbf{Volume measure comparison on BOSS mock data} ---
    \textit{Top panels:}
    Marginal cosmological posteriors from $\Lambda$CDM fits to BOSS-like synthetic data, where $\omega_{\rm b}$ is held fixed. 
    The total effective volume $V_{\rm eff} = 10 \, {\rm Gpc}^3$, is subdivided in one, two, or four skies (see table~\ref{tab:boss_setup}), wherein the number of marginalised nuisance parameters increases from 12, 24, to 48, enhancing, under a flat measure, volume projection effects on the inferred cosmological parameters. 
    Compared to the flat measure (\textit{grey contours}), which exhibits volume projection effects up to $\sim 1\sigma$, non-flat measures shift the posterior mean closer to the true values (\textit{dashed lines}), reducing bias to below $0.05-0.15\sigma$, depending on the prescription used: as a post-debiasing correction (\textit{blue contours}), or as a log-measure weight (\textit{yellow and black dotted contours}). 
    Constraints obtained using a covariance rescaled by $n = 100$ (\textit{red contours}) are also shown. 
    \textit{Bottom panel:}
    Relative mean biases on cosmological parameters for various volume measures for each sky setup. 
    The relative shifts to the truth of 1D marginal posterior means are expressed in units of the posterior standard deviation. 
    }\label{fig:fake_boss}
\end{figure}

\begin{figure}[!ht]
    \centering
    \footnotesize
    \includegraphics[width=0.8\textwidth]{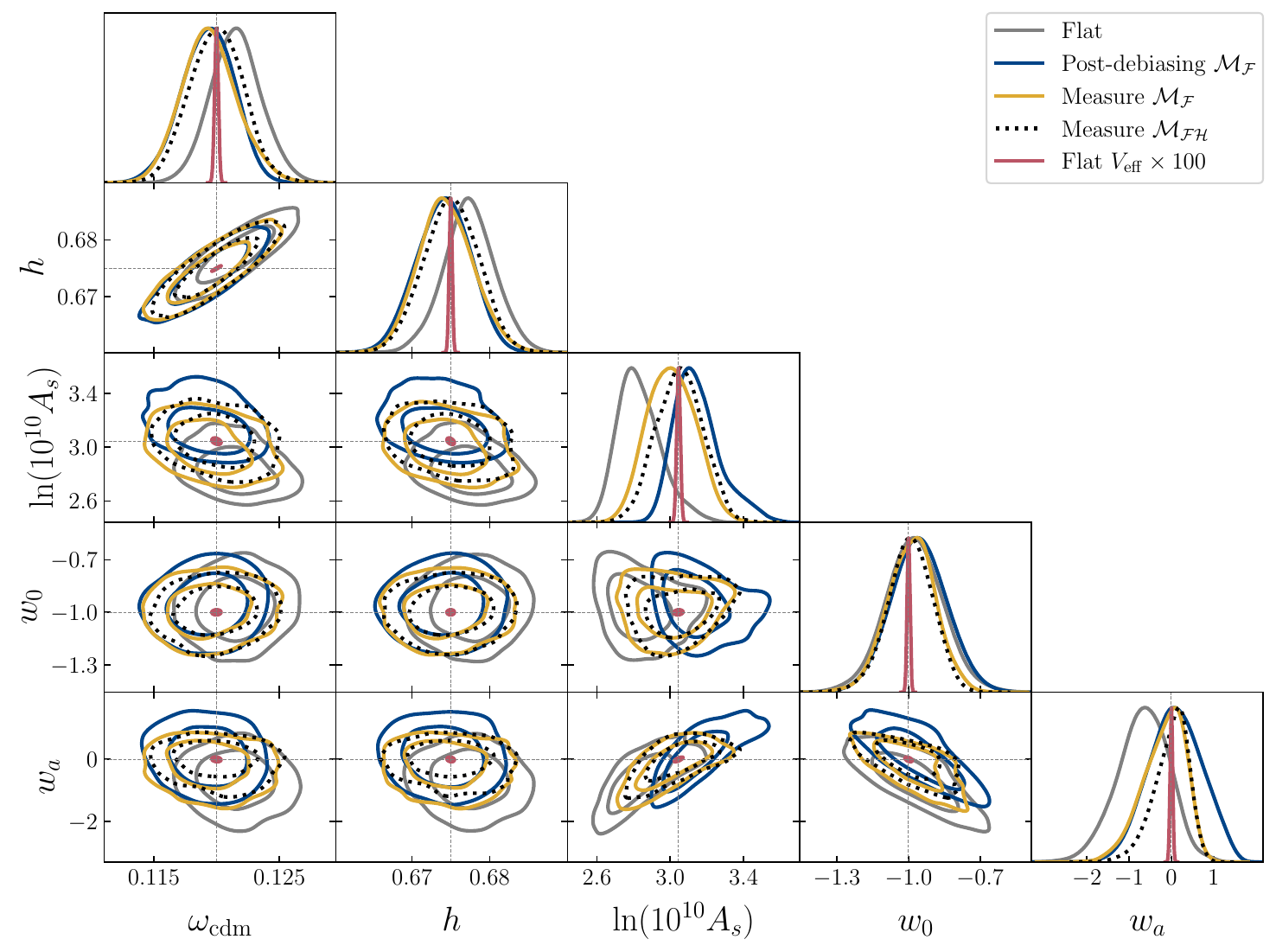}\\
    \vspace{.3cm}
    \begin{tabular}{l l
                  S[table-format=2.2]
                  S[table-format=2.2]
                  S[table-format=2.2]
                  S[table-format=2.2]
                  S[table-format=2.2]}
    \toprule
    \textbf{Model} & \textbf{Volume measure}
      & {$\omega_{\mathrm{cdm}}$}
      & {$h$}
      & {$\ln(10^{10}A_s)$}
      & {$w_0$}
      & {$w_a$} \\
    \midrule
    \multirow{4}{*}{$\Lambda$CDM} 
      & Flat measure                         &  0.68 &  0.57 & -0.92 &  &  \\
      & Post-debiasing $\mathcal{M}_\mathcal{F}$    & -0.30 & -0.20 &  0.26 &  &  \\
      & Measure $\mathcal{M}_\mathcal{F}$           &  0.09 &  0.07 &  0.05 &  &  \\
      & Measure $\mathcal{M}_{\mathcal{F}\mathcal{H}}$         &  0.04 & -0.01 & 0.01 &  & \\
      \multirow{1}{*}{\textit{$V_{\mathrm{eff}}\times100$}} 
       & Flat measure                          & -0.09  & -0.13 &  0.13  &  & \\

    \addlinespace
    
    \multirow{4}{*}{\(w_0w_a\)CDM}
      & Flat measure                         &  0.76 &  0.72 & -1.73 &  0.21 & -1.07 \\
      & Post-debiasing $\mathcal{M}_\mathcal{F}$    & -0.34 & -0.27 &  0.80 &  0.37 &  0.15 \\
      & Measure $\mathcal{M}_\mathcal{F}$           & -0.25 & -0.17 & -0.28 &  0.12 & -0.35 \\
      & Measure $\mathcal{M}_{\mathcal{F}\mathcal{H}}$        &  0.00 &  0.01 &  0.06 &  0.07 & -0.03 \\
    \multirow{1}{*}{\textit{$V_{\mathrm{eff}}\times100$}} 
    & Flat measure                         &  -0.09 &  -0.13 & 0.09 &  0.02 & 0.04 \\
    \bottomrule
  \end{tabular}
    \caption{\textbf{Volume measure comparison on DESI mock data} --- 
    \textit{Top panels:}
    Marginal cosmological posteriors from  $w_0w_a$CDM fits to DESI-like synthetic data, where $\omega_{\rm b}$ and $n_s$ are held fixed. 
    The total effective volume $V_{\rm eff} = 63 \, {\rm Gpc}^3$ is subdivided in seven skies (see table~\ref{tab:desi_y6_setup}), corresponding to a total of 84 marginalised nuisance parameters.  
    Compared to the flat measure (\textit{grey contours}), which exhibits volume projection effects up to the $\sim 1-2 \sigma$ level, non-flat measures shift the posterior mean closer to the true values (\textit{dashed lines}), reducing bias below $\sim 0.8\sigma, \sigma/3$, or $0.1\sigma$ depending on the prescription used: as a post-debiasing correction (\textit{blue contours}), or as a log-measure weight (\textit{yellow and black dotted contours}). 
    Constraints obtained using a covariance rescaled by $n = 100$ (\textit{red contours}) are also shown, illustrating that biases vanish in the large-$n$ limit. 
    \textit{Bottom panel:}
    Relative mean biases on cosmological parameters for various volume measures. 
    The relative shifts to the truth of 1D marginal posterior means are expressed in units of the posterior standard deviation. 
    $\Lambda$CDM results are also shown for comparison. }\label{fig:fake_desi}
\end{figure}

\paragraph{Large-$n$ limit}  
First, we consider the case where the number of repeated samples $n$ becomes large. 
Since we are working with noiseless synthetic data, increasing $n$ is equivalent to multiplying the effective data volume $V_{\rm eff}$ by $n$, or, equivalently, rescaling the covariance as $C \rightarrow n^{-1} C$.  
In figs.~\ref{fig:fake_boss} and~\ref{fig:fake_desi}, we present results for $n = 100$. 
At this data volume, the ground truth is recovered within approximately $0.1\sigma$ for all cosmological parameters. 
This relative bias is significantly smaller than that observed for $n = 1$, which can reach the $1-2\sigma$ level. 
This illustrates that volume projection effects vanish in the large-$n$ limit, as discussed in sec.~\ref{sec:large-n}. 
We now turn to the situation where the available data volume corresponds to actual realistic settings, \textit{i.e.}, $n=1$. 

\paragraph{Large-$N$ enhancement}
Fig.~\ref{fig:fake_boss} shows the marginal cosmological posterior distributions from fitting BOSS mock data. 
We vary all $\Lambda$CDM parameters except $\omega_{\rm b}$, which is held fixed, mimicking a prior from Big Bang Nucleosynthesis experiments. 
The relative shifts of the posterior means with respect to the true fiducial values are also shown. 
Under a flat measure, we observe that the mean bias increases as the number of skies grows (at fixed data volume $V_{\rm eff}$). 
For instance, the mean bias in $\ln(10^{10}A_s)$ increases (in absolute value) from $0.4\sigma$ to $0.9\sigma$ when moving from one to four skies. 
More skies imply more nuisance parameters, increasing the phase-space volume that projects onto the cosmological parameters upon marginalisation. 
Although many EFT parameters are linear, they remain coupled to cosmological parameters, as discussed in sec.~\ref{sec:EFTofLSS}. 
Thus, volume projection effects become increasingly significant as the number of marginalised parameters $N$ grows. 

\paragraph{Post-debiasing}
We next compare the results from the flat measure with those shifted by our post-debiasing correction. 
For the BOSS-like synthetic data, the mean bias on one sky reduces from $\sim \sigma/3$ to below $0.01\sigma$ for all cosmological parameters. 
On four skies, the bias decreases from $\sim 0.5-0.9\sigma$ to below $0.15\sigma$. 
On these setups, the leading-order correction to the mean bias thus suffices to stay within a $\sigma/3$ tolerance. 
Similar conclusions hold for DESI-like data in $\Lambda$CDM as shown in fig.~\ref{fig:fake_desi}, where the bias drops from $\sim 0.6-0.9\sigma$ to below $\lesssim 0.3\sigma$. 
In contrast, for $w_0w_a$CDM, the bias in $\ln(10^{10}A_s)$ only reduces from $1.7\sigma$ to $0.8\sigma$, while other parameters are close to our $\sigma/3$ tolerance after post-debiasing. 
To investigate these discrepancies, we turn to the results obtained with full non-flat measures. 

\paragraph{Non-flat measures}
In all $\Lambda$CDM fits, using the non-flat volume measure $\mathcal{M}_{\mathcal{F}}$ recovers cosmological parameters within $<0.1\sigma$ of the truth. 
For the $w_0w_a$CDM fit to the DESI-like data, the mean bias also falls below or near our $\sim \sigma/3$ tolerance. 
For instance, the bias in $\ln(10^{10}A_s)$ reduces from $1.7\sigma$ to $0.3\sigma$ using $\mathcal{M}_{\mathcal{F}}$, outperforming the post-debiasing correction (which left a residual shift of $0.8\sigma$). 
Two factors can explain this improvement.  
First, the non-flat measure can correct higher-order bias beyond the leading-order one in the Laplace expansion computed in sec.~\ref{sec:efficient}. 
Second, the measure can also adjust the variance and higher moments, affecting the relative mean shift when expressed in units of $\sigma$. 
This is especially visible on the 1D posterior of $\ln(10^{10}A_s)$, where the flat-measure posterior appears more asymmetric than its non-flat counterparts. 
In this work, we have eluded bias in the variance and higher moments, which are higher orders than the terms we have considered. 
Thus, while definitive conclusions remain out of reach, our empirical results suggest that non-flat measures can substantially further reduce relative bias compared to relying solely on the leading-order correction. 
Importantly, we verified against the Fisher matrix that the Cramér-Rao bound remains satisfied, indicating that non-flat measures do not induce unreasonable variance reductions.

\paragraph{$\mathcal{M}_\mathcal{F}$ vs. $\mathcal{M}_\mathcal{H}$}
Finally, we compare the Jeffreys measure $\mathcal{M}_\mathcal{F}$ to the optimal measure $\mathcal{M}_\mathcal{H}$, predicted to correct more bias in the mean. 
Since we work with noiseless synthetic data, we use the hybrid measure $\mathcal{M}_\mathcal{FH}$ in place of $\mathcal{M}_\mathcal{H}$. 
In $\Lambda$CDM, $\mathcal{M}_\mathcal{FH}$ reduces 
the mean bias below $0.05\sigma$ accross all cosmological parameters and survey setups. 
For $w_0w_a$CDM on DESI-like data, the mean bias falls below $0.07\sigma$ under $\mathcal{M}_\mathcal{FH}$, an improvement of up $\sim 0.2-0.3\sigma$ compared to $\mathcal{M}_\mathcal{F}$. 
These appreciable shifts remain modest compared to the much larger mean bias (up to $1.1-1.7\sigma$) observed under the flat measure.  
These results confirm expectations from sections~\ref{sec:Nbias}~and~\ref{sec:EFTofLSS}:
most of the mean bias arises from large-$N$ enhanced volume effects mitigated by the Jeffreys measure, while the optimal measure $\mathcal{M}_\mathcal{H}$ corrects for potential residual secondary bias from nonlinear parameter dependence (\textit{e.g.}, in $b_1, \ln(10^{10}A_s), w_a$).


\section{Conclusions and discussions}\label{sec:conclusions}
In the last decades, inference in cosmology has been dominated by Bayesian techniques. 
As we have seen, ambiguity can arise in reporting credible values for a physical quantity given the posterior distribution. 
It has to do with the way we define and choose our statistical estimators --- this is what we have touched on in this paper. 
Undesirable aspects of Bayesian inference are often attributed to its inevitable dependence on the prior. Even when no particular prior is assumed (as it is a desiderata to measure physical quantities) we are still setting a flat uniform measure in directions of the parameters we choose to sample. Expectation values over the resulting marginal posteriors therefore necessarily depend on how this specific volume measure gets projected. At first sight this dependence might appear arbitrary, given that equivalent parametrisations of a theory lead to different prior volume. As such, one may decide to disregard this as a real issue, thinking that after all we are talking about $\mathcal{O}(1)$ relative biases and, as more data are collected, these will eventually fade away. We provided two objections.

First, in general, it proves useful to gain insights from current data, as physics is constructed upon lessons learnt incrementally. As a matter of fact, it seems unavoidable to be every now and then in situations where naïve estimators are biased to some extent, while interpreting the results remains necessary to orient research in relevant ways.\footnote{It is noteworthy that, across various contexts in physics (see \textit{e.g.}, refs.~\cite{DAgostini:1999gfj,Creminelli:2001ij,Hamann:2007pi,Planck:2013nga,DESI:2024jis}), different communities and generations have independently encountered similar concerns regarding this issue, and particularly when experimental precision was not yet sufficient to render prior dependence negligible.}  
In that regard, precision matters. Our work provides a way to achieve efficient parameter estimation away from the large sample limit, by untangling the ambiguity in the choice of the volume measure.
Second, at a more concrete level, the examples we have seen seem to indicate that the statement ``\textit{Assuming that our model is true, statistical bias in our physical results will fade away as data increases}'' might not be true in general. Perhaps in cosmology we are facing such a situation, as discussed in length in sec.~\ref{sec:EFTofLSS}. 

\paragraph{}
We now summarise our findings as follow and point towards possible future directions:

\begin{itemize}
\item On average, for asymptotic normal distributions, the posterior mode is biased, as evidenced by eq.~\eqref{eq:modebias}. 
The bias cannot be corrected easily as it depends intrinsically on the nominal true values. 
For practical situations, we however note that the mode bias is not large-$N$ enhanced, as it can be compared to eq.~\eqref{eq:largeN}. 

\item On average, for asymptotic normal distributions, the mean is biased, as evidenced by eq.~\eqref{eq:meanbias}. 
Contrary to the mode, the average mean bias can be calculated with the sole knowledge of the mode. 
This is due to a cancellation between two contributions to the average mean bias: the average mode bias and the contribution~\eqref{eq:modebias2mean}. 
The remaining contribution~\eqref{eq:meanbias} depends only on the (biased) mode. 
It would be interesting to understand if this cancellation is accidental or if similar ones hold beyond leading order in the Laplace expansion, especially in higher-$p$ moments. 

\item As shown in sec.~\ref{sec:Nbias}, for an inferred parameter $\theta_\mu$, the average mean bias is enhanced by the number of parameters $N$ that are marginalised when those are coupled to $\theta_\mu$, \textit{i.e.}, the model second derivatives $\partial_{\mu\nu} m$ are non-zero. 
This can lead to arbitrarily large bias ($>1\sigma$) for large $N$, without loss of asymptotic convergence (see however ref.~\cite{McCullagh2} for a strict convergence criterion accounting for both $N$ and $n$).  
In contrast, the self-couplings $\alpha = \partial_{\mu\mu}m$ lead to a parametrically small bias, of relative size $\sim \alpha/\sigma$. 

\item To correct the bias in marginal posteriors and resulting credible intervals, we propose two alternative estimators for the posterior mean.  
The first, that we dub post-debiased mean, consists in shifting the standard posterior mean $\mathbb{E}_{\mathcal{P}}[\theta_\alpha]$ (with expectation value defined with respect to the Lebesgue measure) by the average bias we have estimated, 
\begin{equation}\label{eq:final0}
\mathbb{E}'_{\mathcal{P}}[\theta_\alpha] = \mathbb{E}_{\mathcal{P}}[\theta_\alpha] - b [\mathcal{F}, \pmb {\theta_*}] \ ,
\end{equation}
where $b [\mathcal{F}, \pmb {\theta_*}]$ is given by eq.~\eqref{eq:meanbias} (or eq.~\eqref{eq:meanbias_noiseless} for noiseless synthetic data). 
The second, that relies on defining the expectation value with respect to a non-flat measure, is given by
\begin{equation}\label{eq:final}
\mathbb{E}_{\mathcal{P},\mathcal{M}_{\mathcal{H}}}[\theta_\alpha] = \int \mathcal{M}_\mathcal{H}(\pmb\theta) \, \theta_\alpha \, \mathcal{P}(\pmb\theta |y) \ ,
\end{equation}
where the measure $\mathcal{M}_\mathcal{H}(\pmb\theta)=\mathcal{M}_\mathcal{H}(\pmb\theta|\mathcal{F},\pmb{\theta_*})$ is given by eq.~\eqref{eq:H}. 
As shown in sec.~\ref{sec:estimator}, the two estimators agree in both their mean and variance at leading order (up to $\mathcal{O}(n^{-2})$ corrections) in the Laplace expansion. 
Arguably, defining a measure that depends on the mode $\pmb{\theta_*}$ does not appear problematic to us as the expectation value is defined in any case with respect to the posterior $\mathcal{P}$, and a fortiori on the mode (that lives in $\mathcal{P}$). 

\item Other choices of volume measure can be considered. 
We have shown that, the Jeffreys measure $\mathcal{M}_\mathcal{F}(\pmb \theta) \equiv \sqrt{\det \mathcal{F}(\pmb \theta)}d^N\pmb \theta$ corrects for the large-$N$ enhanced bias. 
The \emph{Jeffreys prior} can not be interpreted as a \emph{prior distribution} in a formal sense (though sometimes referred as \textit{improper} prior distribution). 
Instead, it is well defined as a \emph{prior volume measure}, in the sense that it can be chosen \emph{a priori} to knowledge of $\mathcal{P}$, being reliant only on the Fisher $\mathcal{F}$. 

\item A concerned Bayesian may object against a parameter-dependent measure. 
However, we note that even a flat measure is parameter dependent, in the sense that it depends on the specific parametrisation chosen for the model: a flat measure in one parametrisation will be non-flat in another one. 
In contrast to $\mathcal{M}_\mathcal{H}$ or $\mathcal{M}_\mathcal{F}$, the posterior mean under a flat measure is not invariant under reparametrisation, leading to subjective results that depend on the choice of parametrisation. 

In fact, it is possible to perform a change of variables such that, at least locally, $\mathcal{M_F}$ \textit{appears} flat in the new parametrisation. 
If there exists a repametrisation function $\pmb\phi = h(\pmb\theta) \in \mathbb{R}^N$ such that the Jacobian $J_{\mu\nu}^{-1} = \partial\theta_\mu/\partial\phi_\nu$ satisfies 
\begin{equation}
|\det J^{-1}| =  \sqrt{\det \mathcal{F(\pmb\theta)}} \ ,
\end{equation}
this ensures that (see eq.~\eqref{eq:invariance}),
\begin{equation}
\sqrt{\det \mathcal{F(\pmb\theta)}} \, d^N\pmb \theta = d^N\pmb \phi \ . 
\end{equation}
By decomposing the Fisher via Cholesky around $\pmb \theta_*$ as $\mathcal{F_*} = LL^T$, at least locally, we have $\pmb \phi = L^T (\pmb \theta - \pmb {\theta_*})$. 
Assuming a flat Lebesgue measure in $\pmb\phi$-space is then equivalent to assuming a non-flat Jeffreys measure in $\pmb\theta$-space (at least at leading order in the Taylor expansion of the Jeffreys measure).  
The same line of thoughts obviously applies to $\mathcal{M_H}$. 
This offers an heuristic perspective on volume measures and projection effects: projection effects are minimised in the coordinate system where the curvature encoded by the metric $\mathcal{F}_{\mu\nu}$ appears locally flat. 
In any other coordinate system, the volume measure is non-flat and determined by the Jacobian of the transformation, when it exists. 

\item We have shown in explicit examples in large-scale structure how the bias from volume projection effects on inferred cosmological parameters is reduced drastically when taking expectation values with respect to non-flat measures. 
In most cases, $\mathcal{M}_{\mathcal{F}}$ is found to be enough to recover the true values within a tolerance of $\lesssim \sigma/3$, while $\mathcal{M}_{\mathcal{H}}$ is always superior (reducing the bias below $<0.1\sigma$). 
Empirically, we find that defining all moments of the distributions with respect to non-flat measures (in practice adding a log-measure weight to each log-posterior samples) mitigates better the bias than the post-debiasing correction, in the sense that the relative shift to the truth is reduced. 
Do non-flat measures mitigate higher-order biases beyond the leading-order term considered in this work, such as the $\mathcal{O}(n^{-2})$ bias in the mean or variance? 
Computing these next-to-leading-order biases and investigating in which extent non-flat measures can correct for them would be valuable.

\item In the limit where the expansion parameter $\epsilon \sim \alpha/\sigma$ becomes too large, asymptotic convergence is lost. 
Experimental (synthetic) data shows that for $\epsilon \lesssim 1/3$, where $\epsilon$ can be estimated as the second term of eq.~\eqref{eq:meanbias} (the large-$N$-independent bias) relative to the posterior standard deviation, non-flat measures correct efficiently volume projection bias in inference. 
When $\epsilon \gtrsim 1/3$, we find that the bias is large whatever the volume considered. 
Estimating $\epsilon$ can thus be used a priori as a criteria to understand if a given inference will lead to meaningful results.

\item We have not attempted to look at situations where the truth may lie close to a bound, for which volume projection effects may be important --- positiveness in amplitude parameters such as neutrino mass, abundance fraction of a new species, etc. 
It would be interesting to investigate if non-flat measures can help in mitigating projection effects in these cases, for example starting from a log-measure on the amplitude parameter bounded from below. 

\item On the practical side, we have focused on the analyses of galaxy clustering based on the full shape of the power spectrum, where volume projection effects have attracted significant attention (see, \textit{e.g.}, refs.~\cite{DAmico:2022osl,Simon:2022lde,Donald-McCann:2023kpx,Maus:2023rtr,Maus:2024dzi,DESI:2024jis,Paradiso:2024yqh}). 
In particular, ref.~\cite{Paradiso:2024yqh} introduced a numerical iterative method that reparametrises nuisance parameters so that their subspace becomes orthogonal to that of (\textit{i.e.}, decorrelate from) the cosmological parameters of interest, finding in specific examples that most volume projection effects are mitigated. 
Their reparametrisation is nonlinear, implying that their method can, in principle, correct for contributions beyond the leading-order bias derived in this work (as shown in sec.~\ref{sec:estimator}, the leading-order correction from non-flat measures is linear in $\pmb \theta$). 
It would be interesting to explore the connection between their numerical approach and non-flat measure prescriptions.  

\item Profile likelihoods have recently attracted a lot of attention as an alternative for parameter estimation in cosmology (see \textit{e.g.}, refs.~\cite{Herold:2024enb,Holm:2023laa, Holm:2022kkd,2401.14225,2309.04468,2308.06379,Yeche:2005wn,Planck:2013nga}). 
While offering complementary perspective, they face two practical obstacles: (\textit{i}) they depend on the posterior mode, which is itself a biased estimator, so confidence regions constructed around it can be systematically displaced, and (\textit{ii}) each likelihood point must be obtained through a separate high‑accuracy optimisation, causing the computational cost to scale rapidly with the number of points in the grid search. 
Moreover, confidence intervals in profile likelihood analyses rely on accurate point-estimates of $\chi^2$, whereas upon posterior sampling numerical noise tend to average out. Profiling is then typically prohibitive even for two‑parameter profiles. Our debiased marginal posterior estimation circumvents both issues. 

\item In light of our findings, it would be valuable to revisit other realistic analysis setups where volume projection effects have been observed (see \textit{e.g.},~refs.~\cite{1308.4704,2111.13619,2106.13821}). 
In particular, full-shape analyses incorporating the one-loop bispectrum in the EFTofLSS~\cite{DAmico:2022ukl}, which involve a significantly larger number of EFT parameters, have shown projection effects up to the $2\sigma$ level on BOSS data when uncorrected~\cite{DAmico:2022osl}. 
Ref.~\cite{DAmico:2022osl} employed a linear prior numerically tuned on synthetic data, which should be effectively equivalent to adding a log-measure weight Taylor-expanded at first order as done in eq.~\eqref{eq:Hexpand}. 
Given differentiable likelihoods incorporating the one-loop bispectrum --- which we intend to develop, extending the work of our companion paper~\cite{paper1} --- the analytical methods presented here should yield comparable results in a more straightforward manner.
We plan to re-investigate such analyses with renewed expectations; that is, under a non-flat volume measure.
\end{itemize}

At the very least, the programme developed here quantifies how volume projection biases contaminate marginal statistics, so that one can isolate geometric artefacts in them, and if desired, use our non-flat measure prescription. 
Furthermore, we routinely compare marginalised constraints between experiments, often advertising their credible intervals in isolation. If those intervals are distorted by uncontrolled projection volumes, such comparisons become misleading. 
A transparent bias-corrected construction of these marginal estimates, as proposed here, can aid parameter comparison across experiments. 


\section*{Acknowledgements}
We thank Guido D'Amico, Leonardo Senatore, and especially Peter McCullagh, for discussions and comments on the draft. 
PZ thanks the organisers and the participants of the workshops \textit{Theoretical Modeling of Large-Scale Structure of the Universe} at the Higgs Centre for Theoretical Physics, Edinburgh, and \textit{New Strategies for Extracting Cosmology from Galaxy Surveys - 2nd edition} at the Center for Astrophysics, Sesto, for the inspiring atmospheres that helped renew perspectives and motivate the present work. 
PZ acknowledges support from Fondazione Cariplo under the grant No 2023-1205. 

%
%
%
%

\appendix

\section{Validity of Laplace expansion}\label{app:validity}

In this appendix, we review the assumptions outlined in ref.~\cite{Kass} that a posterior distribution must satisfy for the Laplace expansion of its $p$-moments, as used in this work, to be valid. 
Let $\mathcal{P}(\pmb{\theta} | y)$ be a probability distribution of parameters $\pmb{\theta} \in \mathbb{R}^N$ inferred from the likelihood~\eqref{eq:likelihood}. 
For simplicity we assume a flat prior $\pi(\pmb{\theta}) \equiv 1$. 
If $\mathcal{P}(\pmb{\theta} | y)$ satisfies:

\begin{enumerate}
\item[(\textit{i})] \emph{Smooth local extremal region:} $\log \mathcal{P}$ has a mode (local maxima) $\pmb{\theta_*}$ and is smooth in its vicinity, say a ball $B_\epsilon(\pmb{\theta_*})$ of radius $\epsilon > 0$ centred on $\pmb{\theta_*}$; around this maxima, all derivatives of $\log  \mathcal{P}$ are finite, \emph{i.e.}, for $\pmb \theta \in B_\epsilon(\pmb{\theta_*}) \ , |\partial_{\alpha_1}  \log \mathcal{P}(\pmb{\theta})| \ ,  |\partial_{\alpha_1 \alpha_2} \log  \mathcal{P}(\pmb{\theta})| \ , \dots$ are bounded; the determinant of the Hessian of $\log \mathcal{P}$ evaluated at the mode is positive, \emph{i.e.}, $\det H(\pmb{\theta_*}) > 0$, where $H(\pmb{\theta}) := \partial_{\mu \nu} \log \mathcal{P}(\pmb{\theta})$;
\item[(\textit{ii})] \emph{Exponentially-decaying distant regions:} Distant regions to the mode, say the exterior of a ball $\mathbb{R}^N \setminus B_\delta(\pmb{\theta_*})$ of radius $0 < \delta < \epsilon$, have exponentially decreasing probability in the large $n$ limit, \emph{i.e.}, for $p \in \mathbb{N}$, 
\begin{equation}
\sqrt {|H(\pmb{\theta_*})|} \cdot \int_{\mathbb{R}^N \setminus B_\delta(\pmb{\theta_*})} d^N\pmb \theta \  \theta_{\alpha_{1,\dots,p}}  \ \frac{\mathcal{P}(\pmb{\theta})}{\mathcal{P}(\pmb{\theta_*})} \sim \mathcal{O}(n^{-2}) \ ;
\end{equation}
\end{enumerate}
then the $p$-moments of $\mathcal{P}(\pmb{ \theta} | y)$, eq.~\eqref{eq:moment}, can be expanded around $\pmb{\theta_*}$ following Laplace's method. 
Results up to $\mathcal{O}(n^{-3/2})$ are presented in section~\ref{sec:laplace} and up to $\mathcal{O}(n^{-2})$ (for which the assumptions above are for) in appendix~\ref{app:n-2}. 

\section{Moments at $\mathcal{O}(n^{-2})$}\label{app:n-2}
Here we provide expansions for the posterior distribution, zeroth, first, second, and third, moments, up to $\mathcal{O}(n^{-2})$. 
These can be used as a fast, approximate way to obtain the posterior distribution without sampling, together with data/noise-dependent asymmetric credible intervals.
Expanding the posterior using Laplace method, we get
\begin{align}
& \mathcal{P}(\pmb{\theta}|\mathcal{F}, \pmb{j}) =  \mathcal{P}(\pmb{\theta_*})  \exp \left( -\frac{1}{2}\delta_\mu \mathcal{F}_{\mu\nu} \delta_\nu + j_\mu \delta_\mu \right) \times 
\Bigg\{ 1 + \frac{n^{-1/2}}{2} \left[  j_{\mu;\nu}  \delta_\mu \delta_\nu - \frac{1}{2} \mathcal{F}_{\mu\nu;\rho} \delta_\mu \delta_\nu \delta_\rho \right]  \nonumber \\
& + \frac{n^{-1}}{2} \left[ \frac{1}{3} j_{\mu;\nu\rho}  \delta_\mu \delta_\nu \delta_\rho - \frac{1}{4}\left(\frac{1}{2}\mathcal{F}_{\mu\nu;\rho\sigma}+\frac{1}{3}\mathcal{F}^\mu_{\nu;\rho\sigma}\right)\delta_\mu \delta_\nu \delta_\rho \delta_\sigma \right]  \nonumber \\
& + \frac{n^{-3/2}}{12}  \left[\frac{1}{2}  j_{\mu;\nu\rho\sigma}  \delta_\mu \delta_\nu \delta_\rho  \delta_\sigma - \frac{1}{6}\left( \mathcal{F}_{\mu\nu;\rho\sigma\eta} + \mathcal{F}^\mu_{\nu;\rho\sigma\eta} \right)\delta_\mu \delta_\nu \delta_\rho \delta_\sigma \delta_\eta \right]  \Bigg\} + \dots  \ ,
\end{align}
where all quantities are evaluated at $\pmb{\theta_*}$. 
Here and in the rest of this appendix, $\dots$ refer to $\mathcal{O}\left(n^{-2}\right)$. 
Likewise, 
\begin{align}
f(\pmb{\theta}) = f_* + n^{-1/2} f_{;\mu} \delta_\mu + \frac{n^{-1}}{2} f_{;\mu\nu} \delta_\mu  \delta_\nu + \frac{n^{-3/2}}{3!}  f_{;\mu\nu\rho} \delta_\mu  \delta_\nu \delta_\rho  + \dots \ ,
\end{align}
where $f_* =  f(\pmb{\theta_*})$. 
Thus, 
{\footnotesize
\begin{align}\label{eq:n-2}
& f(\pmb{\theta})\mathcal{P}(\pmb{\theta}|\mathcal{F}, \pmb{j})  = \mathcal{P}(\pmb{\theta_*})  \exp \left( -\frac{1}{2}\delta_\mu \mathcal{F}_{\mu\nu} \delta_\nu + j_\mu \delta_\mu \right) \times \Bigg\{ f_* +  n^{-1/2} \left[f_{;\mu} \delta_\mu + \frac{f_*}{2} j_{\mu;\nu} \delta_\mu \delta_\nu - \frac{f_*}{4}\mathcal{F}_{\mu\nu;\rho} \delta_\mu \delta_\nu \delta_\rho \right]   \nonumber \\
& \qquad + \frac{n^{-1}}{2}  \left[f_{;\mu\nu} \delta_\mu  \delta_\nu +  \left( \frac{f_*}{3}  j_{\mu;\nu\rho} + f_{;\rho}  j_{\mu;\nu} \right) \delta_\mu \delta_\nu \delta_\rho - \frac{1}{2}\left(\frac{f_*}{4}\mathcal{F}_{\mu\nu;\rho\sigma}+\frac{f_*}{6}\mathcal{F}^\mu_{\nu;\rho\sigma} +  f_{;\sigma}\mathcal{F}_{\mu\nu;\rho} \right) \delta_\mu \delta_\nu \delta_\rho \delta_\sigma \right]  \nonumber \\
& \qquad  +  \frac{n^{-3/2}}{6} \left[ f_{;\mu\nu\rho} \delta_\mu  \delta_\nu \delta_\rho + \left(\frac{f_*}{4}j_{\mu;\nu\rho\sigma} + f_{;\sigma}j_{\mu;\nu\rho} + \frac{3f_{;\rho\sigma}}{2}j_{\mu;\nu} \right) \delta_\mu  \delta_\nu \delta_\rho \delta_\sigma   \right. \nonumber \\
& \qquad \qquad \qquad \qquad  \left. - \left( \frac{f_*}{12} \mathcal{F}_{\mu\nu;\rho\sigma\eta} + \frac{f_*}{12}\mathcal{F}^\mu_{\nu;\rho\sigma\eta}  + \frac{3f_{;\eta}}{8}\mathcal{F}_{\mu\nu;\rho\sigma} + \frac{f_{;\eta}}{4}\mathcal{F}^\mu_{\nu;\rho\sigma} + \frac{3f_{;\sigma\eta}}{4} \mathcal{F}_{\mu\nu;\rho} \right) \delta_\mu  \delta_\nu \delta_\rho \delta_\sigma \delta_\eta \right] \Bigg\} + \dots 
\end{align} }
The zeroth, first, and second moments of $\mathcal{P}$ then read
{\footnotesize
\begin{align*}
\mathcal{Z}_\mathcal{P} / G[\pmb{j}]  & =  1 +  \frac{n^{-1/2}}{2} \left[ j_{\mu;\nu} g_{\mu\nu} - \frac{1}{2}\mathcal{F}_{\mu\nu;\rho} g_{\mu\nu\rho}  \right]  + \frac{n^{-1}}{6}  \left[j_{\mu;\nu\rho}  g_{\mu\nu\rho} - \frac{1}{4}\left(\frac{3}{2}\mathcal{F}_{\mu\nu;\rho\sigma} + \mathcal{F}^\mu_{\nu;\rho\sigma} \right)g_{\mu\nu\rho\sigma} \right] 
\nonumber \\
& \qquad + \frac{n^{-3/2}}{24} \left[ j_{\mu;\nu\rho\sigma} g_{\mu\nu\rho\sigma} - \frac{1}{3}\left( \mathcal{F}_{\mu\nu;\rho\sigma\eta} + \mathcal{F}^\mu_{\nu;\rho\sigma\eta} \right) g_{\mu\nu\rho\sigma\eta} \right]   + \dots \ , \\
\tilde{\mathbb{E}}_{\mathcal{P}}[\delta_\alpha]  / G[\pmb{j}] & = n^{-1/2}  g_{\alpha} + \frac{n^{-1}}{2}\left[ j_{\mu;\nu} g_{\alpha\mu\nu} - \frac{1}{2}\mathcal{F}_{\mu\nu;\rho} g_{\alpha\mu\nu\rho} \right] + \frac{n^{-3/2}}{6}\left[j_{\mu;\nu\rho} g_{\alpha\mu\nu\rho} - \frac{1}{4}\left(\frac{3}{2}\mathcal{F}_{\mu\nu;\rho\sigma} +\mathcal{F}^\mu_{\nu;\rho\sigma}\right) g_{\alpha\mu\nu\rho\sigma}\right]  + \dots \ , \\
\tilde{\mathbb{E}}_{\mathcal{P}}[\delta_\alpha\delta_\beta] / G[\pmb{j}] &  =  n^{-1} g_{\alpha\beta} + \frac{n^{-3/2}}{2}\left[ j_{\mu;\nu}g_{\alpha\beta\mu\nu} - \frac{1}{2}\mathcal{F}_{\mu\nu;\rho}g_{\alpha\beta\mu\nu\rho} \right] + \dots \ ,
\end{align*} }
where we use the shorthands $g_{\mu} \equiv \partial \log G / \partial j_\mu$, etc.
Perturbatively inverting the evidence yields
{\footnotesize
\begin{align}
\mathbb{E}_{\mathcal{P}}[\delta_\alpha]  & =  n^{-1/2}  g_{\alpha} + \frac{n^{-1}}{2} \bigg[ j_{\mu;\nu}  \left( g_{\alpha\mu\nu} - g_{\alpha} g_{\mu\nu} \right)  - \frac{1}{2}\mathcal{F}_{\mu\nu;\rho} \left( g_{\alpha\mu\nu\rho} - g_{\alpha} g_{\mu\nu\rho} \right) \bigg]  \nonumber \\
&  + \frac{n^{-3/2}}{6} \bigg[ j_{\mu;\nu\rho} (g_{\alpha\mu\nu\rho}-g_\alpha g_{\mu\nu\rho}) 
- \frac{1}{4}\left(\frac{3}{2}\mathcal{F}_{\mu\nu;\rho\sigma} +\mathcal{F}^\mu_{\nu;\rho\sigma}\right)(g_{\alpha\mu\nu\rho\sigma}- g_\alpha g_{\mu\nu\rho\sigma})  \nonumber \\
&  - \frac{3}{2} j_{\mu;\nu}j_{\rho;\sigma} \, g_{\alpha\mu\nu}g_{\rho\sigma} - \frac{3}{4}\mathcal{F}_{\mu\nu;\rho}\mathcal{F}_{\sigma\eta;\epsilon} g_{\alpha\mu\nu\rho}g_{\sigma\eta\epsilon} + \frac{3}{4} j_{\mu;\nu}\mathcal{F}_{\rho\sigma;\eta} (g_{\mu\nu}g_{\alpha\rho\sigma\eta} + g_{\alpha\mu\nu}g_{\rho\sigma\eta})   \bigg]  + \dots \ , \\
\mathbb{E}_{\mathcal{P}}[\delta_\alpha\delta_\beta]  & = n^{-1} g_{\alpha\beta} + \frac{n^{-3/2}}{2}\left[ j_{\mu;\nu} (g_{\alpha\beta\mu\nu} - g_{\alpha\beta}g_{\mu\nu}  ) - \frac{1}{2} \mathcal{F}_{\mu\nu;\rho} (g_{\alpha\beta\mu\nu\rho} -g_{\alpha\beta}g_{\mu\nu\rho}) \right] + \dots \ , \\
\mathbb{E}_{\mathcal{P}}[\delta_\alpha\delta_\beta\delta_\gamma]  & = n^{-3/2} g_{\alpha\beta\gamma} + \dots \ , 
\end{align} }
where we provide the expression for the third moment for completeness.

%
%
%
%

\section{Diagrammatic representation}\label{sec:feyman}

\begin{figure}[ht!]
\centering
\begin{tikzpicture}[xscale=1.1, yscale=1.2]

\node at (0,2.5) {\( j_{\mu} \)};
\node[fill, circle, minimum size=6pt, inner sep=0pt] (v1) at (0,2.0) {};
\node[draw, minimum size=6pt, inner sep=0pt] (s1) at (0,0.5) {};
\draw[thick] (v1) -- (s1.north);
\node at (0,0.2) {\( \mu \)};

\node at (2,2.5) {\( j_{\mu\nu} \)};
\node[fill, circle, minimum size=6pt, inner sep=0pt] (v2) at (2,2.0) {};
\node[draw, minimum size=6pt, inner sep=0pt] (s2a) at (1.5,0.5) {};
\node[draw, minimum size=6pt, inner sep=0pt] (s2b) at (2.5,0.5) {};
\draw[thick] (v2) -- (s2a.north);
\draw[thick] (v2) -- (s2b.north);
\node at (1.5,0.2) {\( \mu \)};
\node at (2.5,0.2) {\( \nu \)};

\node at (4,2.5) {\( j_{\mu\nu\rho} \)};
\node[fill, circle, minimum size=6pt, inner sep=0pt] (v3) at (4,2.0) {};
\node[draw, minimum size=6pt, inner sep=0pt] (s3a) at (3.4,0.5) {};
\node[draw, minimum size=6pt, inner sep=0pt] (s3b) at (4.0,0.5) {};
\node[draw, minimum size=6pt, inner sep=0pt] (s3c) at (4.6,0.5) {};
\draw[thick] (v3) -- (s3a.north);
\draw[thick] (v3) -- (s3b.north);
\draw[thick] (v3) -- (s3c.north);
\node at (3.4,0.2) {\( \mu \)};
\node at (4.0,0.2) {\( \nu \)};
\node at (4.6,0.2) {\( \rho \)};

\node at (5.5,1.4) {\Large$\dots$};

\node[draw, minimum size=6pt, inner sep=0pt] (sq1) at (6.8,0.5) {};
\node[draw, minimum size=6pt, inner sep=0pt] (sq2) at (8.0,0.5) {};
\draw[thick] (sq1) -- (sq2);
\node at (7.4,1.2) {\( \mathcal{F}^{-1}_{\mu\nu} \)};
\node at (6.8,0.2) {\( \mu \)};
\node at (8.0,0.2) {\( \nu \)};

\node[fill, circle, minimum size=6pt, inner sep=0pt] (vc1) at (9.3,2.0) {};
\node[fill, circle, minimum size=6pt, inner sep=0pt] (vc2) at (10.5,2.0) {};
\draw[thick] (vc1) -- node[above=4pt] {\( \mathcal{F}_{\mu\nu} \)} (vc2);

\node[draw, minimum size=6pt, inner sep=0pt] (sq3) at (9.3,0.5) {};
\node[draw, minimum size=6pt, inner sep=0pt] (sq4) at (10.5,0.5) {};
\draw[thick] (vc1) -- (sq3.north);
\draw[thick] (vc2) -- (sq4.north);
\node at (9.3,0.2) {\( \mu \)};
\node at (10.5,0.2) {\( \nu \)};

\node[fill, circle, minimum size=6pt, inner sep=0pt] (vf1) at (11.7,2.0) {};
\node[draw, minimum size=6pt, inner sep=0pt] (sf1) at (11.7,0.5) {};
\draw[thick] (vf1) -- (sf1.north);
\node at (11.7,0.2) {\( \mu \)};

\node[fill, circle, minimum size=6pt, inner sep=0pt] (vf2) at (12.9,2.0) {};
\node[draw, minimum size=6pt, inner sep=0pt] (sf2a) at (12.4,0.5) {};
\node[draw, minimum size=6pt, inner sep=0pt] (sf2b) at (13.4,0.5) {};
\draw[thick] (vf2) -- (sf2a.north);
\draw[thick] (vf2) -- (sf2b.north);
\node at (12.4,0.2) {\( \nu \)};
\node at (13.4,0.2) {\( \rho \)};
\draw[thick] (vf1) -- node[above=4pt] {\( \mathcal{F}^{\mu}_{\nu\rho} \)} (vf2);

\node at (14.3,1.4) {\Large$\dots$};

\end{tikzpicture}
\caption{Diagrammatic representation of source vertices and Fisher propagators}
\label{fig:rule0}
\end{figure}
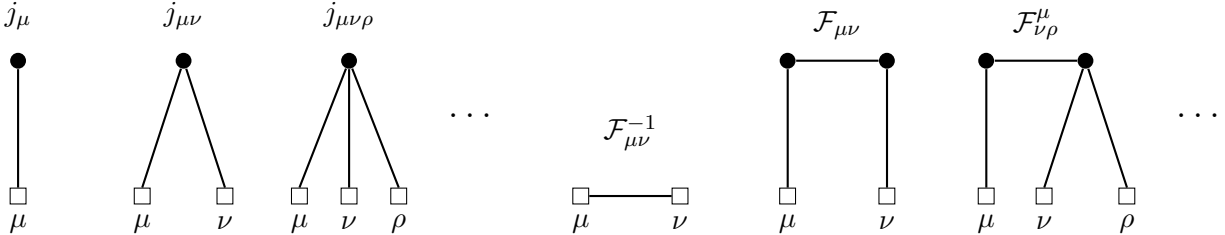

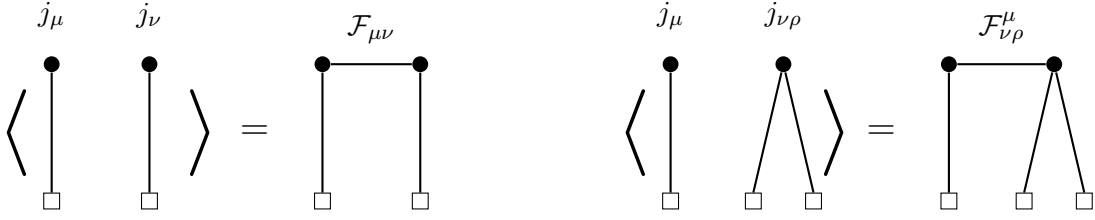
\begin{figure}[ht!]
\centering

\begin{minipage}{0.48\textwidth}
\centering
\begin{tikzpicture}[xscale=1.0, yscale=1.3]

\node at (-0.4,0.3) {\scalebox{3}{$\langle$}};
\node at (2.1,0.3) {\scalebox{3}{$\rangle$}};

\node at (0.1,1.5) {\( j_\mu \)};
\node[fill, circle, minimum size=6pt, inner sep=0pt] (v1) at (0.1,1.0) {};
\node[draw, minimum size=6pt, inner sep=0pt] (s1) at (0.1,-0.4) {};
\draw[thick] (v1) -- (s1.north);

\node at (1.4,1.5) {\( j_\nu \)};
\node[fill, circle, minimum size=6pt, inner sep=0pt] (v2) at (1.4,1.0) {};
\node[draw, minimum size=6pt, inner sep=0pt] (s2) at (1.4,-0.4) {};
\draw[thick] (v2) -- (s2.north);

\node at (2.8,0.3) {\Large$=$};

\node[fill, circle, minimum size=6pt, inner sep=0pt] (v3) at (3.7,1.0) {};
\node[draw, minimum size=6pt, inner sep=0pt] (s3) at (3.7,-0.4) {};
\draw[thick] (v3) -- (s3.north);

\node[fill, circle, minimum size=6pt, inner sep=0pt] (v4) at (5.0,1.0) {};
\node[draw, minimum size=6pt, inner sep=0pt] (s4) at (5.0,-0.4) {};
\draw[thick] (v4) -- (s4.north);

\draw[thick] (v3) -- node[above=4pt] {\( \mathcal{F}_{\mu\nu} \)} (v4);

\end{tikzpicture}
\end{minipage}
\hfill
\begin{minipage}{0.48\textwidth}
\centering
\begin{tikzpicture}[xscale=1.0, yscale=1.3]

\node at (-0.4,0.3) {\scalebox{3}{$\langle$}};
\node at (2.3,0.3) {\scalebox{3}{$\rangle$}};

\node at (0.1,1.5) {\( j_\mu \)};
\node[fill, circle, minimum size=6pt, inner sep=0pt] (v1) at (0.1,1.0) {};
\node[draw, minimum size=6pt, inner sep=0pt] (s1) at (0.1,-0.4) {};
\draw[thick] (v1) -- (s1.north);

\node at (1.6,1.5) {\( j_{\nu\rho} \)};
\node[fill, circle, minimum size=6pt, inner sep=0pt] (v2) at (1.6,1.0) {};
\node[draw, minimum size=6pt, inner sep=0pt] (s2a) at (1.2,-0.4) {};
\node[draw, minimum size=6pt, inner sep=0pt] (s2b) at (2.0,-0.4) {};
\draw[thick] (v2) -- (s2a.north);
\draw[thick] (v2) -- (s2b.north);

\node at (2.9,0.3) {\Large$=$};

\node[fill, circle, minimum size=6pt, inner sep=0pt] (v3) at (3.8,1.0) {};
\node[draw, minimum size=6pt, inner sep=0pt] (s3) at (3.8,-0.4) {};
\draw[thick] (v3) -- (s3.north);

\node[fill, circle, minimum size=6pt, inner sep=0pt] (v4) at (5.2,1.0) {};
\node[draw, minimum size=6pt, inner sep=0pt] (s4a) at (4.8,-0.4) {};
\node[draw, minimum size=6pt, inner sep=0pt] (s4b) at (5.6,-0.4) {};
\draw[thick] (v4) -- (s4a.north);
\draw[thick] (v4) -- (s4b.north);

\draw[thick] (v3) -- node[above=4pt] {\( \mathcal{F}^{\mu}_{\nu\rho} \)} (v4);

\end{tikzpicture}
\end{minipage}
\caption{Master rule 1 --- $\braket{ j_\mu j_\nu } = \mathcal{F}_{\mu\nu}$, $\braket{ j_\mu j_{\nu\rho} } = \mathcal{F}^\mu_{\nu\rho}$, $\dots$ }
\label{fig:rule1}
\end{figure}

\begin{figure}[ht!]
\centering

\begin{tikzpicture}[xscale=1.0, yscale=1.3]


\node at (-0.6,0.4) {\scalebox{3}{$\langle$}};
\node at (4.2,0.4) {\scalebox{3}{$\rangle$}};

\node[draw, minimum size=6pt, inner sep=0pt] (sMuL) at (0,0) {};
\node at (0,-0.4) {\( \mu \)};
\node[draw, minimum size=6pt, inner sep=0pt] (sNuL) at (1.2,0) {};
\node[fill, circle, minimum size=6pt, inner sep=0pt] (cNuL) at (1.2,1.4) {};
\draw[thick] (sMuL) -- (sNuL);
\draw[thick] (sNuL) -- (cNuL);
\node at (1.2,1.8) {\( g_\mu \)};

\node[draw, minimum size=6pt, inner sep=0pt] (sMuR) at (2.4,0) {};
\node[draw, minimum size=6pt, inner sep=0pt] (sNuR) at (3.6,0) {};
\node at (3.6,-0.4) {\( \nu \)};
\node[fill, circle, minimum size=6pt, inner sep=0pt] (cMuR) at (2.4,1.4) {};
\draw[thick] (sMuR) -- (sNuR);
\draw[thick] (sMuR) -- (cMuR);
\node at (2.4,1.8) {\( g_\nu \)};

\node at (4.9,0.4) {\Large$=$};


\node[draw, minimum size=6pt, inner sep=0pt] (sMuL2) at (5.8,0) {};
\node at (5.8,-0.4) {\( \mu \)};
\node[draw, minimum size=6pt, inner sep=0pt] (sMidL2) at (7.0,0) {};
\node[fill, circle, minimum size=6pt, inner sep=0pt] (cTopL2) at (7.0,1.4) {};
\draw[thick] (sMuL2) -- (sMidL2);
\draw[thick] (sMidL2) -- (cTopL2);
\node at (6.4,0.3) {\( \mathcal{F}^{-1}_{\mu\rho} \)};

\node[draw, minimum size=6pt, inner sep=0pt] (sMidR2) at (8.2,0) {};
\node[draw, minimum size=6pt, inner sep=0pt] (sNuR2) at (9.4,0) {};
\node at (9.4,-0.4) {\( \nu \)};
\node[fill, circle, minimum size=6pt, inner sep=0pt] (cTopR2) at (8.2,1.4) {};
\draw[thick] (sMidR2) -- (sNuR2);
\draw[thick] (sMidR2) -- (cTopR2);
\node at (8.8,0.3) {\( \mathcal{F}^{-1}_{\sigma\nu} \)};

\draw[thick] (cTopL2) -- (cTopR2);
\node at (7.6,1.7) {\( \mathcal{F}_{\rho\sigma} \)};

\node at (10.1,0.4) {\Large$\equiv$};


\node[draw, minimum size=6pt, inner sep=0pt] (sFinalMu) at (11.0,0) {};
\node at (11.0,-0.4) {\( \mu \)};

\node[draw, minimum size=6pt, inner sep=0pt] (sFinalNu) at (12.2,0) {};
\node at (12.2,-0.4) {\( \nu \)};

\draw[thick] (sFinalMu) -- (sFinalNu);
\node at (11.6,0.3) {\( \mathcal{F}^{-1}_{\mu\nu} \)};

\end{tikzpicture}

\caption{Master rule 2 --- $\braket{ g_\mu g_\nu } = \mathcal{F}^{-1}_{\mu\nu}$, where $g_\mu = \mathcal{F}^{-1}_{\mu\rho} j_\rho$ }
\label{fig:rule2}
\end{figure}
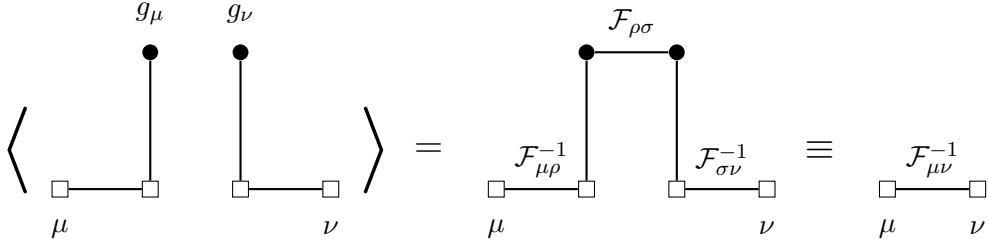

In this appendix, we provide in figs.~\ref{fig:rule0},~\ref{fig:rule1},~and~\ref{fig:rule2} the diagrammatic representations and rules to compute average $p$-moments of asymptotic normal posterior distributions. 

\section{Analytic marginalisation}\label{app:marg}

Let $m$ be a model described by parameters $\pmb{\theta} = \lbrace \pmb{\Omega}, \pmb{\psi} \rbrace$ where $\pmb{\psi}$ are linear parameters and $\pmb{\Omega}$ the others. 
Up to an irrelevant constant, the log-posterior given the likelihood~\eqref{eq:likelihood} together with a Gaussian prior $\pi(\pmb \theta)$ reads
\begin{align}
\ln \mathcal{P}(\pmb{\theta}) & = -\frac{1}{2}(m(\pmb{\theta})-y) \cdot C^{-1} \cdot (m(\pmb{\theta})-y) + \ln \pi(\pmb{\theta}) \ , \label{eq:logp} \\
\ln \pi(\pmb{\theta}) & = -\frac{1}{2}(\psi_\alpha - \hat \psi_\alpha) \cdot \mathcal{C}^{-1}_{\alpha \beta} \cdot (\psi_\beta - \hat \psi_\beta) + \ln \Pi(\pmb{\Omega}) \ ,
\end{align}
where the prior $\pi$ consists in a multivariate Gaussian centred on $\pmb{\hat{\psi}}$ with covariance $\mathcal{C}$ for the linear parameters $\pmb{\psi}$ and a general prior $\Pi$ for the other parameters. 
We now regroup the linear parameters $\pmb \psi$ that we refer using $(\alpha, \beta, \dots)$ indices when running in vector/tensor quantities, as opposed to $(\mu, \nu, \dots)$ indices that run on all parameters $\pmb \theta$, or $(i,j \dots)$ indices running only on $\pmb \Omega$. 
The $\log$-posterior becomes
\begin{equation}\label{eq:quadratic}
\ln \mathcal{P}(\pmb{\Omega}, \pmb{\psi}) = -\frac{1}{2}\psi_\alpha \mathcal{F}_{\alpha\beta}(\pmb{\Omega}) \psi_\beta + j_\alpha(\pmb{\Omega})  \psi_\alpha + \ln \mathcal{P}|_{\pmb{\psi} = 0} \ ,
\end{equation}
where 
\begin{align}
\mathcal{F}_{\alpha\beta}(\pmb{\Omega})  & = \partial_\alpha m \cdot C^{-1} \cdot \partial_\beta m + \mathcal{C}^{-1}_{\alpha\beta} \ , \\
j_\alpha(\pmb{\Omega})  & = -\partial_\alpha m \cdot C^{-1} \cdot (m|_{\pmb{\psi} = 0} - y) + \mathcal{C}^{-1}_{\alpha\beta} \hat \psi_\beta \ .
\end{align}
Given that the log-posterior~\eqref{eq:quadratic} is explicitly quadratic in $\pmb \psi$, we can marginalising over $\pmb{\psi}$ using properties of Gaussian integrals, yielding
\begin{equation}\label{eq:marg}
\ln \tilde{\mathcal{P}}(\pmb{\Omega}) = \frac{1}{2}j_\alpha \mathcal{F}^{-1}_{\alpha\beta} j_\beta + \ln \mathcal{P}|_{\pmb{\psi} = 0} -\frac{1}{2}\ln \det |\mathcal{F}_{\alpha \beta}| \ . 
\end{equation}
For a prior centred on $\hat \psi_\alpha = 0$, eq.~\eqref{eq:marg} is sometimes re-written in the form of a usual Gaussian likelihood on the subspace $\pmb \Omega$~\cite{1003.1136,Bridle:2001zv}. 
Up to a measure, it reads
\begin{equation}
\ln \tilde{\mathcal{P}}(\pmb{\Omega}) = -\frac{1}{2}(m|_{\pmb{\psi}=0}-y) \cdot \tilde{C}^{-1} \cdot (m|_{\pmb{\psi}=0}-y) + \ln \Pi(\pmb \Omega) \ , 
\end{equation}
where, defining $\mathcal{U}_\alpha \equiv C^{-1} \cdot \partial_\alpha m$, the precision matrix $\tilde{C}^{-1}$ that depends on $\pmb \Omega$ is
\begin{equation}\label{eq:newC}
\tilde{C}^{-1} = C^{-1} - \mathcal{U}_\alpha \mathcal{F}_{\alpha \beta}^{-1} \mathcal{U}_\beta \ .
\end{equation}

\paragraph{Integration measure}
We now show that weighting posterior distribution with measure $\sqrt{\det |\mathcal{F}_{\mu\nu}|}$ is equivalent to the following. 
$\mathcal{F}_{\mu\nu}$ is the full Fisher matrix of $\pmb \theta$, while $\mathcal{F}_{ij}$ and $\mathcal{F}_{\alpha\beta}$ denote the submatrices associated to $\pmb \Omega$ and $\pmb \psi$, respectively, with  $\mathcal{F}_{i\alpha}$ their cross-matrix. 
To proceed, we can make use of the Shur complements formula, 
\begin{equation}
\det \mathcal{F}_{\mu\nu} = \det \mathcal{F}_{\alpha\beta} \cdot \det F_{ij} \ , \qquad F_{ij} := \mathcal{F}_{ij} - \mathcal{F}_{i\alpha} \mathcal{F}_{\alpha\beta}^{-1} \mathcal{F}_{\beta j} \ , 
\end{equation}
or equivalently
\begin{equation}\label{eq:shurlogdet}
\frac{1}{2}\ln \det \mathcal{F}_{\mu\nu} = \frac{1}{2}\ln \det F_{ij} + \frac{1}{2}\ln \det \mathcal{F}_{\alpha\beta} \ .
\end{equation}
Adding the second term of~\eqref{eq:shurlogdet} to the log-posterior is equivalent to dropping the log-determinant from eq.~\eqref{eq:marg}, yielding $\ln \bar{\mathcal{P}}$.  
Then, one can then show that $F_{ij}$ is the average Hessian of $\ln \bar{\mathcal{P}}$.\footnote{We thank Guido d'Amico for pointing to us this fact. } 
Thus in practice what we are left with simply amounts to sample $\bar{\mathcal{P}}$ weighted by half of the log-determinant of its Fisher information, as how it is done for posteriors without analytic marginalisation.


 \bibliographystyle{JHEP}
 \small
\bibliography{references}

 \end{document}